\documentclass[journal, lettersize]{IEEEtran}

\usepackage{background}
\usepackage{amsmath,amsfonts}
\usepackage{url}
\usepackage{cite}
\usepackage{booktabs}
\usepackage{caption}
\usepackage{comment}
\usepackage{xcolor}
\usepackage{xspace}
\usepackage{listings}
\usepackage{bm}
\usepackage{array}
\usepackage{longtable}
\usepackage{supertabular}
\usepackage{multirow}
\usepackage{array}
\usepackage[many]{tcolorbox}
\usepackage{colortbl}
\usepackage{paralist}
\usepackage{soul}
\usepackage{pifont}
\newcommand{\cmark}{\ding{51}}%
\newcommand{\xmark}{\ding{55}}%

\definecolor{darkred}{HTML}{860000}
\definecolor{darkteal}{HTML}{005959}
\definecolor{darkpurple}{HTML}{590059}
\definecolor{darkgrey}{HTML}{434343}

\backgroundsetup{
  scale=2.5,
  color=gray,
  opacity=0.2,
  angle=45,
  contents={\textbf{ACCEPTED}}
}

\newcolumntype{?}{!{\vrule width 1pt}}

\newcommand{\completeECount}{24\xspace}

\newcommand{\tool}{ChatGPT\xspace}
\newcommand{\codefont}[1]{\footnotesize{\texttt{#1}}\normalsize}

\newcommand{\gpt}{GPT-4.0\xspace}
\newcommand{\lama}{Llama3.3-70b\xspace}
\newcommand{\gemini}{Gemini-2.5-pro-preview\xspace}
\newcommand{\claud}{Claude-3.7-sonnet\xspace}
\newcommand{\deepseek}{Deepseek-chat-v3\xspace}

\newcommand{\gptcomp}{35\xspace}
\newcommand{\lamacomp}{36\xspace}
\newcommand{\geminicomp}{34\xspace}
\newcommand{\claudcomp}{35\xspace}
\newcommand{\deepcomp}{32\xspace}

\newcommand{\gptfix}{15*\xspace}
\newcommand{\lamafix}{16*\xspace}
\newcommand{\geminifix}{9*\xspace}
\newcommand{\claudfix}{6*\xspace}
\newcommand{\deepfix}{13*\xspace}

\newcommand{\lamademo}{23\xspace}
\newcommand{\geminidemo}{18\xspace}
\newcommand{\clauddemo}{17\xspace}
\newcommand{\deepdemo}{20\xspace}
\newcommand{\gptdemo}{23\xspace}

\newcommand{\lamademofix}{10*\xspace}
\newcommand{\geminidemofix}{5*\xspace}
\newcommand{\clauddemofix}{3*\xspace}
\newcommand{\deepdemofix}{7*\xspace}
\newcommand{\gptdemofix}{7*\xspace}

\newcommand{\totalClient}{49\xspace}
\newcommand{\totalVulClient}{49\xspace}
\newcommand{\totalGPTClient}{46\xspace}

\newcommand{\totalLib}{25\xspace}
\newcommand{\totalVul}{29\xspace}
\newcommand{\VulFilter}{45\xspace}
\newcommand{\totalCompTest}{38\xspace}
\newcommand{\runnable}{26\xspace}
\newcommand{\minorfix}{12\xspace}
\newcommand{\notCompTest}{11\xspace}

\newcommand{\ying}[1]{\textcolor{black}{#1}}
\newcommand{\noCVE}{36\xspace}
\newcommand{\noM}{39\xspace}

\newcommand{\noC}{15\xspace}
\newcommand{\noExaTest}{30\xspace}

\newcommand{\tse}[1]{\textcolor{black}{#1}}

\newcolumntype{L}[1]{>{\raggedright\let\newline\\\arraybackslash\hspace{0pt}}m{#1}}
\newcolumntype{C}[1]{>{\centering\let\newline\\\arraybackslash\hspace{0pt}}m{#1}}
\newcolumntype{R}[1]{>{\raggedleft\let\newline\\\arraybackslash\hspace{0pt}}m{#1}}

\lstdefinestyle{javastyle}{
  language=Java,
  basicstyle=\scriptsize\ttfamily,
  numbers=left,
  numberstyle=\tiny,
  numbersep=5pt,
  frame=lines,
  breaklines=true,
  framesep=1mm,
  showstringspaces=false,
  columns=flexible
}
\usepackage{colortbl}%
\newtcolorbox{mybox}[2][]{text width=0.95\linewidth,fontupper=\footnotesize,
fonttitle=\ttfamily\scriptsize, colbacktitle=darkgrey,enhanced,
attach boxed title to top left={yshift=-2mm,xshift=3mm},
boxed title style={sharp corners},top=4pt,bottom=2pt,left=2pt,right=2pt,
  title=#2,colback=white}

\def\BibTeX{{\rm B\kern-.05em{\sc i\kern-.025em b}\kern-.08em
    T\kern-.1667em\lower.7ex\hbox{E}\kern-.125emX}}
\begin{document}

\title{How Can \tool Support Human Security Testers to Help Mitigate Supply Chain Attacks?} 



\author{Ying Zhang,~\IEEEmembership{Member,~IEEE,}
        Wenjia Song,~\IEEEmembership{}
        Zhengjie Ji,~\IEEEmembership{Student Member,~IEEE,}
        Danfeng (Daphne) Yao,~\IEEEmembership{Fellow,~IEEE,}
        and Na Meng,~\IEEEmembership{Member,~IEEE}
\IEEEcompsocitemizethanks{The work was partially supported by NSF-1845446, NSF-1929701, ONR N00014-22-1-2057, and CCI-PE6GLBAE.}
\IEEEcompsocitemizethanks{Ying Zhang is with the Department of Computer Science, Wake Forest University, Winston-Salem, NC 27109. \protect E-mail: ying.zhang@wfu.edu.}
\IEEEcompsocitemizethanks{Wenjia Song, Zhengjie Ji, Danfeng (Daphne) Yao, and Na Meng are with the Department of Computer Science, Virginia Tech, Blacksburg, VA 24060. \protect 
E-mail: \{wenjia7, zhengjie, danfeng, nm8247\}@vt.edu.}}

\markboth{Journal of \LaTeX\ Class Files,~Vol.~18, No.~9, September~2020}{Systematic Assessment of a Large Language Model for Generating Dependency Exploits}

\maketitle
\begin{abstract}
Developers often build software on top of third-party libraries (Libs) to improve programmer productivity and software quality. The libraries may contain vulnerabilities exploitable by hackers to attack the applications (Apps) built on top of them. Such attacks are known as software \textbf{supply chain attacks}, the documented number of which has increased by 600\% since 2021. Researchers and developers created tools to mitigate such attacks, by scanning the library dependencies of Apps, identifying the usage of vulnerable library versions, and suggesting secure alternatives to vulnerable dependencies. However, recent studies show that many developers do not trust the reports by these tools; they need code or evidence to demonstrate how library vulnerabilities lead to security exploits, in order to assess vulnerability severity and modification necessity. Unfortunately, manually crafting demos of application-specific attacks is challenging and time-consuming, and there is insufficient tool support to automate that procedure.

To help developers enhance software security, in this study, we systematically explored the usage of a large language model (LLM)---ChatGPT-4.0---to generate \textbf{security tests}, which unit tests demonstrate how vulnerable library dependencies facilitate the supply chain attacks to given \emph{Apps}. 
In our exploration, we defined prompt templates to take in the various vulnerability-relevant information we manually collected, and generated prompts from those templates to query \tool for security test generation. 
We found that \tool-generated tests demonstrated \completeECount evidence or proof of vulnerability for \totalVulClient Apps.
\tse{To assess the consistency of test generation, we also evaluated another five state-of-the-art LLMs. All the models generated security tests for at least 17 cases that successfully demonstrate the vulnerabilities.} 
{We filed six reports for the newly revealed vulnerabilities \tse{in Apps}, and got four Common Vulnerability Entries (CVEs) assigned}. Our use of ChatGPT outperformed two state-of-the-art security test generators (TRANSFER and SIEGE), by generating a lot more tests and achieving more attacks. 
Our research will shed light on new research in security test generation.
\end{abstract}


\section{Introduction}\label{sec:intro}

\begin{figure}[ht]
\centering
\vspace{-1em}
\includegraphics[width=.75\linewidth]{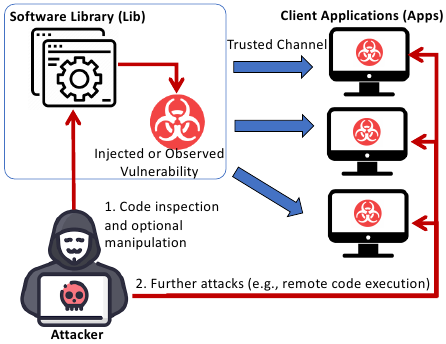}
\vspace{-.5em}
\caption{The threat model of supply chain attacks}\label{fig:threat-model}
\vspace{-.5em}
\end{figure}

\IEEEPARstart{S}oftware development relies on open-source external dependencies and third-party APIs to accelerate development. Developers often integrate these APIs without fully vetting their security vulnerabilities~\cite{meng2018secure, vu2020towards}. 
Recent studies showed that many APIs contain known security flaws~\cite{garrett2019detecting, wu2023understanding}. When software applications invoke vulnerable APIs without properly validating their security properties, vulnerabilities can be propagated from APIs to those applications~\cite{ladisa2022towards, supply-chain-attack}.
As shown in Fig.~\ref{fig:threat-model}, attackers can inject or identify exploitable vulnerabilities in third-party libraries through either (1) contributing code to open-source libraries, (2) directly inspecting the code of the open-source libraries, or (3) consulting publicly available vulnerability databases. They can further exploit these vulnerabilities to propagate attacks through software supply chains and compromise applications built on these libraries.
\tse{Supply chain attacks (e.g., attack through the Log4Shell vulnerability~\cite{log4shell}) have grown more than 600\% and caused 12,000 incidents~\cite{supply-chain-attack-trend} since 2021.} Open Web Application Security Project (OWASP)~\cite{owasp-top-10-2022} listed ``vulnerable and outdated software components'' as the sixth top vulnerability.


To mitigate supply chain attacks, people have created tools to identify vulnerable library dependencies in software applications~\cite{dependencychecker,snyk-test,dependabot,gammaray,auditjs,retirejs,ponta2020detection,Rahkema2022}, 
and even suggest fixes for those vulnerabilities~\cite{snyk-fix,dependabot,Nguyen2020}.  
For instance, snyk-test~\cite{snyk-test} and npm-audit~\cite{npm-audit} are CLI commands that scan JavaScript (JS) applications for their package dependencies, compare those packages against the package lists in predefined vulnerability databases (e.g., CVE), and report a vulnerability for each found match. However, none of these tools demonstrate how identified vulnerabilities translate to concrete exploits in the developer's specific App context (e.g., denial of service)~\cite{Kabir2022,zhang2022automatic}. For instance, Zhang et al.~\cite{zhang2022automatic} sent vulnerability reports together with patching suggestions to developers. Some developers did not trust the reports or their security implications; they demanded \textit{proof-of-concept attacks} to demonstrate the security exploit. Such delayed updates can leave real vulnerabilities unaddressed, putting applications at continued security risk.

 To help developers assess the impact of reported vulnerabilities and prioritize their fixing process, this paper presents our novel research on generating security tests using LLM, for software applications (Apps) with vulnerable library dependencies (Libs).
 Specifically, if an App calls a vulnerable library API, we generate a prompt for ChatGPT using the following information: (1) the API name, (2) the non-private method $M$ inside App that (in)directly calls that vulnerable API, (3) the Java class defining $M$, (4) the Lib test that shows \textbf{evidence or proof of vulnerability (PoV)}, and (5) the vulnerability entry ID (e.g., CVE entry ID). Here, the PoV test executes Lib in specialized ways to show behavioral differences between the vulnerable and patched versions of Lib.
 
With that prompt, we queried \tool to create a security test for App, which test mimics Lib test to show proof of vulnerability in App. The generated test executes the App 
 to (i) propagate vulnerabilities from Libs to Apps via calls of the API, and (ii) trigger abnormal behaviors of the App, such as throwing errors or 
becoming unresponsive to customers' normal requests. When developers run security tests generated in such a way, they can observe vulnerability propagation paths, foresee the serious consequences due to hackers' successful attacks, assess the severity levels, and better decide whether to address those reported vulnerabilities.

The biggest challenge is ensuring that the generated tests (1) execute those vulnerable APIs called by Apps, (2) trigger any problematic behaviors of Apps, and (3) fail when reported vulnerabilities are not fixed. 
Iannone et al.~\cite{iannone2021toward} and Kang et al.~\cite{kang2022test}  created tools to generate tests using  
EvoSuite~\cite{fraser2011evosuite}, the widely used test generation tool. 
 Unfortunately, both tools fail to generate security tests in many cases. They often spend much time producing irrelevant tests but cannot synthesize the specialized test inputs, code, or oracle.

 {Our initial experience with ChatGPT showed 
 its great potential in generating code to satisfy software requirements. 
 Additionally, many vulnerable open-source libraries were recorded in publicly available security databases~\cite{githubadvisories, snyk,booth2013national,hat2007red, vulnerabilities2005common, martin2011cwe}. These databases catalog disclosed software vulnerabilities; they detail the nature of each vulnerability along with the affected library versions. Such a comprehensive documentation can provide a useful context, where we explored the usage of \tool in generating security tests. We investigated the following research questions (RQs) and observed interesting phenomena:  
 


\textbf{RQ1:} \emph{How effectively does \tool generate security tests?} We created a dataset to include (1) \totalLib Libs and (2) \totalClient Apps, with each App depending on a vulnerable library version. For each App, we offered \tool a prompt to describe the  vulnerability, App context, and a security test from Lib showing that the patched and vulnerable versions differ. With the prompt, we asked \tool to generate a test for the App by mimicking the given test. We found that \tool generated tests for all \totalClient Apps, \completeECount of which are security tests that successfully demonstrated evidence or proof of vulnerability (PoV). 

\textbf{RQ2:} 
\emph{How does \tool's security test generation performance differ given various types of prompts?}
By changing the default design of our prompt template, we fed \tool with different subsets of the descriptive information in the above-mentioned prompts (see RQ1). We observed that all information elements provided important guidance to \tool, while the security test from Lib was the most important. Without a security test from Lib provided, none of the generated tests by \tool could successfully demonstrate PoV.

\textbf{RQ3:} \emph{How does \tool compare with existing tools of security test generation?} We applied \tool and two state-of-the-art tools~\cite{iannone2021toward,kang2022test} to the same datasets. 
Surprisingly, \tool outperformed  both tools: it was always able to generate tests, and those tests achieved much higher success rates in realizing PoV.

\textbf{RQ4:} \emph{How does \tool work with few-shot prompting?} We further explored few-shot prompting by offering \tool one or three examples of test-generation tasks. Interestingly, one-shot prompts led to worse results than the default zero-shot prompting, but three-shot prompts led to better results.
}

\textbf{RQ5:} \emph{How effectively do different LLMs generate security tests?} We explored PoV test generation capabilities across both closed-source and open-source LLMs with the default prompt setting. Interestingly, closed-source models (\gpt, \claud, \gemini) demonstrated higher initial test compilability than open-source alternatives (\lama, \deepseek). \tse{While \lama achieved the highest number of successful PoV demonstrations (23)}, each model exhibited unique vulnerability demonstration capabilities.

In summary, this paper makes the following contributions:
\begin{itemize}
    \item \textbf{New LLM experimental methodology.} We explored using a large language model for security test generation. We designed novel prompt templates that take in the PoV-related information that we manually collected, evaluated different ways of using ChatGPT, and compared ChatGPT with state-of-the-art security-test generators: TRANSFER~\cite{iannone2021toward} and SIEGE~\cite{kang2022test}.
    \item \textbf{New characterization of LLM capabilities.} We observed \tool to work effectively, given prompts that cover the relevant domain knowledge.     
    \tool outperformed state-of-the-art tools that leverage complex program analysis and genetic programming to generate tests. With zero-shot prompts, \tool successfully generated \completeECount security tests for \totalClient Apps.
    \item \textbf{New software security contributions.} Some of ChatGPT-generated tests revealed new vulnerabilities, and we published four CVEs: CVE-2023-31441, CVE-2023-37760, CVE-2023-37761, and 
CVE-2023-43151. 
    \item \textbf{New dataset.} We created a dataset of real-world \totalVulClient $\langle Lib, App\rangle$ pairs, which covers vulnerabilities from four big categories.\url{https://github.com/ying-selab/TSE-2025}.
    
\end{itemize}


\section{A Motivating Example}\label{sec:example}

To facilitate discussion, here we introduce a concrete example to show how vulnerabilities in Libs incur security attacks. Bouncy Castle (BC) is a collection of Java APIs used in cryptography~\cite{bouncy-castle}. According to CVE-2020-28052~\cite{cve-2020-28052}, its releases 1.65 and 1.66 have a vulnerability: \codefont{OpenBSDBCrypt.checkPassword(String bcryptString, char[] password)} improperly implements password-checking logic, allowing wrong passwords to be accepted as valid ones.  Listing~\ref{lst:example-exploit} shows the security test defined by a BC version later than 1.66, which demonstrates PoV. Ideally, if a BC version has no vulnerability, the first assertion (line 9) succeeds as the first parameter \codefont{tokenString} was derived from the password \codefont{test-token}; the second assertion (line 12) succeeds as the first parameter \codefont{tokenString} was not from \codefont{wrong-token}. However, BC 1.65 and BC 1.66 fail the second assertion, as the invalid password \codefont{wrong-token} is wrongly considered to match \codefont{tokenString}. Such a security test demonstrates the problematic behaviors of vulnerable library versions, and implies the potential of security exploits (e.g., sending in wrong passwords to pass identity authentication). 

\begin{figure}[b]
\caption{A Lib test to
show PoV in Lib
}
\label{lst:example-exploit}\vspace{-2em}
\begin{lstlisting}[style=javastyle]
public void performTest() throws Exception {
  ... ...
  int costFactor = 4;
  SecureRandom random = new SecureRandom();
  salt = new byte[16];
  for (int i = 0; i < 1000; i++) {
    random.nextBytes(salt);
    final String tokenString = OpenBSDBCrypt.generate("test-token".toCharArray(), salt, costFactor);    
    isTrue(OpenBSDBCrypt.checkPassword(tokenString, "test-token".toCharArray()));
    /* A safe BC version passes the following assertion; a vulnerable one fails the assertion, as it considers unmatched passwords to match. */
    isTrue(!OpenBSDBCrypt.checkPassword(tokenString, "wrong-token".toCharArray()));
  } }
\end{lstlisting}
\vspace{-.5em}
\end{figure}



Although Listing~\ref{lst:example-exploit} shows a PoV of the library, it does not show how Apps built on top of vulnerable BC versions can behave abnormally or get attacked. Existing vulnerability detectors can report such vulnerable API calls in Apps~\cite{rahaman2019cryptoguard,kruger2017cognicrypt,Zhang2022,zhang2022automatic}. However, as they do not show how vulnerable API calls introduce vulnerabilities or induce supply chain attacks to Apps, App developers are reluctant to revise Lib usage accordingly.

In this project, we explore to mitigate supply chain attacks by generating security tests to simulate hackers' potential ways of using Apps maliciously. Our approach is using \tool to generate security tests for Apps to mimic Lib tests. 
\textbf{Our goal is to investigate how effectively \tool generates App-specific security tests, when it is given relevant information about the Lib vulnerabilities, App context, and exemplar tests.} Thus, we assume some tools or domain experts to detect potential vulnerabilities
and provide relevant data 
to ChatGPT for test generation. Once a test is generated successfully, domain experts or developers
can run the App with that test, observe the problematic program behaviors themselves, assess the security implications of supply chain attacks, and decide whether to eliminate vulnerabilities by upgrading library versions or replacing API calls.


\section{Methodology}\label{sec:method}

We conducted an empirical study on ChatGPT's capability of generating PoVs for Apps that depend on vulnerable Libs and call vulnerable APIs. 
We chose to experiment with \tool-4.0, 
because our preliminary work shows that ChatGPT has the best performance across the different LLMs as shown in Section IV-F.

Our study has three phases: dataset construction, prompt design, and result validation. Phase I (Section~\ref{sec:dataset}) collects known vulnerabilities in Libs, Lib tests showing PoV, and Apps that can be affected by their calls to vulnerable APIs. Phase II  (Section~\ref{sec:prompts}) adopts the information collected by Phase I, formulates a variety of prompts for individual $\langle Lib, App\rangle$ pairs, and sends prompts to \tool. These prompts ask \tool to leverage all information provided, to generate security tests for Apps. 
Phase III (Section~\ref{sec:validation}) gathers all outputs by \tool, assesses the quality of generated tests, and evaluates \tool's capability accordingly.

\subsection{Phase I: Dataset Construction}\label{sec:dataset}
While existing datasets~\cite{kang2022test, iannone2021toward} offer potential benchmarks for evaluating security test generation, we found them inadequate for our specific assessment of ChatGPT.
One of the datasets is not quite usable while the other is unrepresentative, so we decided to create a new dataset. Specifically,  
although the dataset used to evaluate TRANSFER~\cite{kang2022test} covers 22 Libs and 42 Apps, 21 of the Apps could not easily run to demonstrate vulnerability PoV due to (1) project removal from GitHub, (2) compilation or dependency issues, (3) missing vulnerable API calls, and (4) unconvincing vulnerabilities in test files that are excluded from software deliverables.
The dataset used to evaluate SIEGE~\cite{iannone2021toward} includes 11 Libs and 11 Apps, where Apps are handcrafted toy projects instead of real-world programs. In order to systematically evaluate the test generated by ChatGPT, we took two steps to create a dataset:  
(1) finding vulnerabilities with exemplar PoV tests, and (2) getting vulnerable Libs as well as dependent client Apps.


\subsubsection{Locating Vulnerabilities with Exemplar PoV Tests}
\label{sec:step1} Vulnerabilities sparsely exist in software libraries.
To efficiently locate vulnerabilities, we started with 
the datasets mentioned by prior work~\cite{ponta2019msr,kang2022test} and initiated our exploration with 628 entries. 
Each entry is linked to a CVE entry or JIRA issue, describing a vulnerable Lib and a GitHub repository showing both the vulnerable and patched versions of Lib. 

Our initial step involved automatically filtering the commit history to identify commits adding test files. The first and fifth authors performed an in-depth analysis to identify the vulnerable API for each of these filtered commits. As vulnerability descriptions may not always precisely pinpoint vulnerable library APIs, the authors spend time understanding the program context and commit details to accurately identify these APIs. They consider the APIs is vulnerable if it (1) is mentioned or implied by the vulnerability description of CVE or JIRA entry and gets revised, (2) directly or indirectly calls the described vulnerable API, (3) is invoked by the described API and is the root cause for the described vulnerability, or (4) shares the same root cause with the described API (i.e., they both call the same root-cause vulnerable method).
To determine which API is the root cause, the first and fifth authors discussed each case together, cross-validating their findings to ensure consistency and reliability.
To extract the PoV test,  
\tse{we} further chose vulnerability entries based on two criteria:  

\begin{itemize}
\item[(a)] Exemplar Security Test: Lib has at least one JUnit test from the patched version, to demonstrate behavioral differences between the vulnerable and patched versions.
\item[(b)] Successful Execution: The patched version of Lib should run smoothly with the security test, requiring no extra manual effort for bug fixing or software (re)configuration. 
\end{itemize} 
Criterion (a) ensures the 
exploitability of confirmed vulnerabilities. Namely, if no Lib test is available to demonstrate PoV, it is hard for us to manually craft and justify the ground truth of test generation. 
Criteria (b) ensures that we can run the security test defined for Lib, to observe the behavioral differences between vulnerable and patched versions. 
In our study, we implemented the two criteria as filters to refine initial vulnerability datasets.
The filters separately removed 427 and 156 entries, 
leaving \VulFilter for further processing. 

\subsubsection{Collecting Libs and Apps for Vulnerabilities}\label{sec:step2}

For each library identified in the {\VulFilter} refined entries (see Section~\ref{sec:step1}), 
\tse{We automatically crawled GitHub using the GitHub API to find client applications depending on these libraries.}  We use the library or package name as keywords, filtering for projects whose dependency versions fall within the vulnerable range, and limiting our crawling to the first 10 pages of results, given the large volume of projects returned. The first, second, and third authors then manually inspected the code of these projects to verify actual usage of the vulnerable functionality (beyond mere dependency inclusion), with the goal of identifying up to four client applications per library that satisfy both criteria \tse{(c) and (d)}:
\begin{itemize}
\item[(c)] At least one non-private Java method (not test) in App directly or indirectly calls vulnerable API(s) in Lib, and gets impacted by the library vulnerability.
\item[(d)] App compiles and runs successfully.
\end{itemize}
Criterion (c) ensures the feasibility of creating PoV tests. 
Basically, if users craft malicious input values to feed certain public or protected method(s) in App (i.e., the callers of vulnerable APIs), they can run vulnerable APIs in malicious ways and thus realize attacks. Criterion (d) ensures that we can check the correctness of tool-generated tests via compilation and program execution.  
\begin{table*}
\caption{The library vulnerabilities and client applications included in our dataset}
\label{tab:lib-data}
\vspace{-.5em}
\scriptsize
\begin{tabular}{R{.05\textwidth}|L{.11\textwidth}| L{.09\textwidth}| L{.1\textwidth}|L{.49\textwidth}|R{.03\textwidth}} 
\toprule
\textbf{Category}& \textbf{Vulnerability Entry ID} &\textbf{Library} &\textbf{Affected Library Versions} &\textbf{Vulnerable API(s) \& Potential Exploit}&\textbf{\# of Apps} \\ \toprule
&CVE-2017-7957 (CWE-20) &XStream~\cite{xstream}&[, 1.4.9] &{\tt XStream.fromXML(...)} mishandles attempts to create an instance of the primitive type ``{\tt void}'' during unmarshalling, leading to a remote application crash, i.e., denial of service (\textbf{DoS}). &2 
\\ \cline{2-6}
&CVE-2018-1000873 (CWE-20)&Jackson-Modules-
Java8~\cite{jackson-modules-java8}&[, 2.9.8)& 
{\tt ObjectMapper.readValue(...)} triggers DoS when it deserializes a very large decimal value to time. & 3
\\ \cline{2-6}
&CVE-2018-11761 (CWE-611)&Apache Tika~\cite{tika}&[0.1, 1.18]&{\tt SAXParser.parse(...)} was not configured to limit entity expansion, and thus could lead to DoS.&1
\\ \cline{2-6}
&CVE-2018-12418 (CWE-835)& Junrar~\cite{junrar} &[,1.0.1)& 
The {\tt Archive} constructor gets into an infinite loop when handling corrupt RAR files. &1
\\ \cline{2-6}
&CVE-2018-1274 (CWE-770) & Spring Data Commons~\cite{spring-data-commons} & [1.13, 1.13.10], [2.0, 2.0.5] &
{\tt PropertyPath.from(...)} allocates resources without limits, and thus can cause DoS due to its consumption of CPU and memory. &1
\\ \cline{2-6}
&CVE-2019-10093 (CWE-770) &Apache Tika&[1.19, 1.21] 
&{\tt Parser.parse(...)} enables a carefully crafted 2003ml or 2006ml file to consume all available SAXParsers in the pool.
&1
\\ \cline{2-6}
Denial of Service (13)&CVE-2019-12402 (CWE-835) & Apache Commons Compress~\cite{apache-commons-compress}&[1.15, 1.18]&Malicious inputs to 
 {\tt ZipArchiveOutputStream.putArchiveEntry(...)} or {\tt ZipEncoding.encode(...)} can cause infinite loops. &1\\ \cline{2-6} 
&CVE-2020-28491 (CWE-770)& Jackson Dataformat: CBOR~\cite{jackson-dataformats-binary}&[, 2.11.4), (2.12.0-rc1, 2.12.1) & {\tt ObjectMapper.createParser(...)} allocates resources without limits; it can cause {\tt java.lang.OutOfMemoryError}. &1\\ \cline{2-6}
&CVE-2021-27568 (CWE-754) &Json-smart~\cite{json-smart-v1,json-smart-v2}&v1:[, 1.3.2), v2: [, 2.3.1), [2.4, 2.4.1)&{{\tt JSONParser.parse(...)} throws an uncaught exception, which can cause an application crash or expose sensitive information. }&2\\\cline{2-6}
&CVE-2021-30468 (CWE-835) &Apache CXF~\cite{apache-cxf}&[, 3.3.11), [3.4.0, 3.4.4)&Malicious inputs to {\tt JsonMapObjectReaderWriter.fromJson(...)} 
or {\tt JsonMapObjectReaderWriter.fromJsonToJsonObject(...)} 
can result in an infinite loop.&1\\ \cline{2-6}
&CVE-2022-45688 (CWE-787) &JSON-java(i.e., hutool-json)~\cite{hutool-json} &[, 20230227)
 &Malicious inputs to {\tt XML.toJSONObject(...)} or {\tt JSONML.toJSONObject(...)} can trigger DoS.&3\\ \cline{2-6}
&TwelveMonkeys-595&TwelveMonkeys~\cite{twelveMonkeys}&[0, 3.6.4)&A corrupt JPEG file to {\tt ImageReader.read(...)} can cause DoS.&2 (1*)
\\ \cline{2-6}
&Zip4j-263&Zip4j~\cite{zip4j}&[0, 2.7.0)&The {\tt ZipFile(...)} constructor can take in a null File reference, which later produces a null pointer exception. &2 \\ \hline

&CVE-2018-1002200 (CWE-22) &Plexus Archiver~\cite{plexus-archiver} &[,3.6.0)& 
{\tt UnArchiver.extract(...)}, {\tt ZipUnArchiver.extract(...)}, and {\tt TarGZipUnArchiver.extract(...)} allow attackers to write to arbitrary files via ``../'' in an archive entry (\textbf{Zip Slip}). &
2
\\ \cline{2-6}
&CVE-2018-1002201 (CWE-22)& ZT Zip~\cite{zt-zip}&[, 1.13)& {\tt ZipUtil.unpack(...)} allows attackers to write to arbitrary files via archive extraction (Zip Slip).&1  
\\ \cline{2-6}
Directory&CVE-2018-19859 (CWE-22) & OpenRefine~\cite{OpenRefine}& [, 3.2-beta) &{\tt ImportingUtilities.allocateFile(...)} allows arbitrary file write via archive extraction (Zip Slip). &1(1*)
 \\ \cline{2-6}
Traversal (5)
&CVE-2021-29425 (CWE-20) &Apache Commons IO~\cite{apache-commons-io}&[, 2.7)&{\tt FileNameUtils.normalize(...)} enables \textbf{directory traversal}, which provides access to files beyond the target file location.&3\\\cline{2-6}
&HTTPCLIENT-1803&Apache HttpClient&[,4.5.3)&
The {\tt URIBuider} constructor, {\tt URIBuilder.setHost(...)}, {\tt URIBuilder.build(...)}, and {\tt URIBuilder.toString(...)} can result in directory traversal.  &1  \\ \hline

Remote Code&CVE-2017-7525 (CWE-22) & Jackson Databind~\cite{jackson-databind}& [, 2.6.7.1) [2.7.0, 2.7.9.1) [2.8.0, 2.8.9)&A deserialization flaw in the library allows maliciously crafted inputs to {\tt ObjectMapper.readValue(...)} to trigger remote code execution. &2 \\ \cline{2-6}
Execution (4)&CVE-2020-26217 (CWE-78) &XStream&[, 1.4.14)& {Malicious inputs to {\tt XStream.fromXML(...)} allow attackers to run arbitrary shell commands.}& 3 \\ \cline{2-6}
&CVE-2021-23899 (CWE-611) &OWASP JSON Sanitizer&[,1.2.2)&{\tt JsonSanitizer.sanitize(...)} may allow hackers to inject arbitrary code into embedding documents.&1\\ \cline{2-6}
&CVE-2022-25845 (CWE-502) &Fastjson~\cite{fastjson}&[, 1.2.83)&{\tt JSON.parseObject(...)} may deserialize untrusted data, allowing hackers to attack remote servers.&2\\ \hline

&CODEC-134 &Apache Commons Codec~\cite{apache-commons-codec} & [, 1.13) &Malicious inputs to {\tt Base64.decodeBase64(...)} or {\tt Base64.decode(...)} can realize
\textbf{covert channel}~\cite{Zander2007}, which creates a capability of transferring data between processes that should not communicate &3 \\ \cline{2-6}
&CVE-2018-1000632 (CWE-91) & Dom4j~\cite{dom4j} & [, 2.1.1)& {Malicious inputs to {\tt DocumentHelper.createElement(...)} or {\tt Branch.addElement(...)} can result in \textbf{XML injection}, which tampers with XML documents.}&  2
 \\ \cline{2-6}
&CVE-2020-13956  &Apache HttpClient~\cite{httpclient}&[, 4.5.13,), [5.0.0, 5.0.3)&Malicious inputs to {\tt CloseableHttpClient.execute(...)} or {\tt URIUtils.extractHost(...)} trigger \textbf{Blind Server-Side Request Forgery (SSRF)}, which attack induces an application to issue a back-end HTTP request to a supplied URL, but the response from the back-end request is not returned to the application's front-end response.&1 \\\cline{2-6}
Others (7)&CVE-2020-13973 (CWE-20) &OWASP JSON Sanitizer~\cite{json-sanitizer}& [,1.2.1)&{\tt JsonSanitizer.sanitize(...)} does not properly escape disallowed characters, and thus facilitates \textbf{cross-site scripting (XSS)}, which enables the browser to unknowingly execute malicious script on the client side and perform actions that are otherwise blocked by the browser's Same Origin Policy. &1 \\\cline{2-6}
&CVE-2020-28052 (CWE-20)&Bouncy Castle &1.65, 1.66&{\tt OpenBSDBCrypt.checkPassword(...)} improperly verifies passwords, allowing wrong ones to be accepted as valid ones.& 2 (1*)\\ \cline{2-6}
&CVE-2020-5408 (CWE-20) &Spring Security~\cite{spring-security}&[4.2.0, 4.2.16), [5.0.0, 5.0.16), [5.1.0, 5.1.10), [5.2.0, 5.2.4), [5.3.0, 5.3.2)& {\tt BCryptPasswordEncoder.encode(...)} presents cryptographic weakness, which may allow hackers to decrypt encrypted messages via a dictionary attack.&2 (2*)\\\cline{2-6}
&CVE-2023-34454 (CWE-20) &snappy-java~\cite{snappy-java}&[, 1.1.10.1)&{\tt Snappy.compress(...)} improperly validates array length, and may cause Access Violation errors. &1\\ 
\bottomrule
\multicolumn{5}{l}{* indicates the number of clients with injected vulnerable dependencies.}\\
 \bottomrule 
\end{tabular}
\end{table*}
\normalsize
This filtering process removed 16 from the \VulFilter entries mentioned above, because we found no client project to satisfy both criteria for those entries.
As shown in Table~\ref{tab:lib-data}, our dataset includes \totalVul vulnerability entries, corresponding to \totalLib unique Libs. 
These libraries cover various domains, such as data processing (e.g., Apache Commons Codec~\cite{apache-commons-codec}), 
web development (e.g., Apache CXF~\cite{apache-cxf}), and 
security (e.g., Spring Security~\cite{spring-security}). 
Most Libs have a single vulnerability (e.g., Dom4j~\cite{dom4j}), while a few have multiple (e.g., XStream~\cite{xstream}). 
In Table~\ref{tab:lib-data}, 
we identified four major categories among the \totalVul vulnerabilities: denial of service, directory traversal, remote code execution, and others. 
\textbf{Affected Library Versions} shows the vulnerable library versions described by each CVE entry or JIRA issue.  \textbf{Vulnerable API(s) \& Potential Exploit} shows the vulnerable APIs and security consequences we summarized by inspecting all relevant data. 

According to our experience, it is very challenging to identify a sufficient number of Apps satisfying (c)--(d) for any vulnerable library. To conduct a representative empirical study with sufficient data points, we extensively explored the version history of the retrieved Apps even though they did not depend on vulnerable library versions in the current version. 
Specifically for each found project \emph{App} satisfying criteria (c)--(d), we examined the version history to determine whether any earlier version, denoted as $App_i$, depends on a vulnerable library version. 
If $App_i$ exists, we checked out $App_i$ to prepare for \tool usage. Otherwise, 
we manually revised the dependency version. For instance, OpenRefine~\cite{OpenRefine} is a library whose versions before 3.2-beta suffer from CVE-2018-19859. However, we only found one client project for it, which depends on a safe version of OpenRefine (i.e., 3.3).
To make sure that this client is still usable in our study, we downgraded the library dependency in the configuration file to 3.1 without further modification. This strategy reflects a common real-world scenario where Apps rely on outdated dependencies due to delayed patching. We do not inject handcrafted vulnerable logic into Apps, as the approach risks introducing artificial or irrelevant behaviors and was therefore avoided. By contrast, our revision preserved the application code logic and only altered the dependency version to formulate realistic conditions.
Our manual revision of dependencies does not compromise the validity of our research, as we fairly compared all approaches of security test generation on the same set of client Apps, no matter whether their vulnerable dependencies are real or injected. 
 The last column in Table~\ref{tab:lib-data} shows the number of clients we included for each Lib. There are five projects 
 with injected vulnerable dependencies, all of which are marked with asterisks (*). At the end of Phase I, our dataset consists of all relevant information for \totalVulClient $\langle App, Lib\rangle$ pairs.

Our newly created dataset overlaps with TRANSFER's dataset~\cite{kang2022test} by sharing 17 Apps in common, while sharing 0 App with SIEGE's dataset~\cite{iannone2021toward}. This fact implies that our dataset is very different from existing ones; it can enrich the knowledge body of exploitable vulnerabilities in Apps. Furthermore, we classified the \totalClient functions-under-test in our dataset into 3 categories: 
\begin{itemize}
\item C1: Functions directly calling vulnerable APIs and sharing the same parameter lists with called APIs. For instance, if a client function $p(int, int)$ is defined to call API $q(int, int)$, the client function belongs to C1.
\item C2: Functions directly calling vulnerable APIs but defining different parameter lists from the called APIs. For instance, if a client function $p(long)$ is defined to call API $q(int, int)$, then the client function belongs to C2. 
\item C3: Functions indirectly calling vulnerable APIs. For instance, if a client public function $p()$ is defined to call API $q()$ in the following manner: $p()\rightarrow r()\rightarrow q()$, where $r()$ is a private client function directly calling $q()$, then $p()$ belongs to C3. 
\end{itemize}
C1--C3 separately cover 20, 20, and 9 functions. 
We hypothesize the complexity comparison of test-generation tasks among these categories to be C1 $<$ C2 $<$ C3, mainly because the less commonality is shared between functions-to-test and vulnerable APIs, the more difficult it is for \tool to generate tests.

\subsection{Phase II: Prompt Design}\label{sec:prompts}

With the information collected in Phase I, we formulated prompts and asked \tool to generate tests. This section first introduces the information elements extracted from our dataset for each $\langle App, Lib\rangle$ pair (Section~\ref{sec:elements}). It then introduces our various ways of constructing \tool prompts using those elements (Sections~\ref{sec:default} and~\ref{sec:variants}).

\subsubsection{Information Elements Extracted}
\label{sec:elements} For each $\langle App, Lib\rangle$ pair, we extracted seven elements for later prompt creation.

\begin{itemize}
\item[\textbf{(i)}] A GitHub project of the client application App.  
\item[\textbf{(ii)}] A vulnerable version of Lib on which App depends.   
\item[\textbf{(iii)}] The vulnerable Lib API(s) called by App.  
\item[\textbf{(iv)}] The non-private method $M$ inside App that directly or indirectly calls vulnerable APIs.
\item[\textbf{(v)}] The Java class C defining method $M$.  
\item[\textbf{(vi)}] The Lib test T (i.e., a Java method). If it indirectly calls vulnerable APIs, the definition of all methods standing between T and APIs is also included. 
\item[\textbf{(vii)}] The vulnerability entry ID (i.e., CVE or JIRA entry ID).
\end{itemize}
Elements (i)--(ii) are essential to represent an $\langle App, Lib\rangle$ pair; (iii) and (vii) describe the library vulnerability; (vi) shows an exemplar PoV test for Lib, and (iv)--(v) present relevant context in App. These seven elements cover all necessary information we can recognize to describe a generation task of the PoV test for App.

\begin{figure}
\centering
\begin{mybox}
    {\textbf{\textit{Prompt Design}}}
Write a JUnit test for function: \textit{[function-name]} in "CLIENT CODE``. The test should verify whether the function \textit{[function-name]} is affected by the vulnerability: \textit{[vulnerability-ID]}, or affected by vulnerable APIs: \textit{[vulnerable-API-list]}. Mimic the test function "TEST FUNCTION:`` \textit{[test-function-name]} I provide to generate the input in your test. For the provided code "CLIENT CODE``, mock the classes and functions that are not Java SE APIs and have no definition provided.
\begin{verbatim}
TEST FUNCTION:
``` Java
[exemplar-test-impl]
```

CLIENT CODE: 
``` Java
[client-code-impl]
```
\end{verbatim}
\end{mybox}
\vspace{-1em}
\caption{Our default prompt template}
\label{fig:template}
\vspace{-1.5em}
\end{figure}

\subsubsection{The Design of Our Default Prompt Template}\label{sec:default} 
We believe that given more comprehensive prompts, \tool is likely to generate better tests. Therefore, we designed a default prompt template to cover all seven elements mentioned above. 
As shown in 
Fig.~\ref{fig:template}, the template has seven variables or customizable parameters to accept data for each task-specifying prompt. In more detail,  
variable \emph{[function-name]} is the name of $M$ (see \textbf{(iv)});  \emph{[vulnerability-ID]} refers to \textbf{(vii)}, \emph{[vulnerable-API-List]} refers to \textbf{(iii)}; \emph{[test-function-name]} is the name of example test (see \textbf{(vi)});  \emph{[exemplar-test-impl]} refers to \textbf{(vi)}; and \emph{[client-code-impl]} refers to \textbf{(v)}.

Overall, the template
asks \tool to generate a JUnit test for a function in App, so that (1) the test  verifies whether that App function is affected by a known vulnerability or by specified vulnerable APIs, and (2) the test mimics the exemplar test to generate malicious inputs. 
With this template, we generated \totalClient prompts. 

\begin{figure}[b]
\caption{A GitHub project defines a Java file to call {\tt OpenBSDBCrypt.checkPassword(String, char[])}}
\label{lst:call}\vspace{-2em}
\begin{lstlisting}[style=javastyle]
public class BcryptPasswordHashFunction implements PasswordHashFunction {
  ...
  // check(...) calls the vulnerable API, so it can be affected by CVE-2020-28052.
  @Override  
  public boolean check(String passwordHash, String password) {
    return OpenBSDBCrypt.checkPassword(passwordHash, password.toCharArray());
  } ... }
\end{lstlisting}
\vspace{-.5em}
\end{figure}

For the motivating example described in Section~\ref{sec:example}, we actually found a GitHub project depending on Bouncy Castle. As shown in Listing~\ref{lst:call}, the project defines a method \codefont{BcryptPasswordHashFunction.check(...)} to directly call vulnerable API \codefont{OpenBSDBCrypt.checkPassword(...)}.  Although both the API and its caller take in two parameters, the caller method \codefont{check(...)} has to convert its second parameter \codefont{password} before calling that API. To generate a security test for \codefont{check(...)}, we formulated the prompt shown in Fig.~\ref{fig:prompt-example} by customizing our template. Fig.~\ref{fig:prompt-example} only shows partial code of the example test and client class to simplify our presentation, while the actual prompt we sent to \tool includes the complete code. 

Given the prompt, \tool successfully generated an executable security test as requested. Listing~\ref{lst:generate} shows a brief version of the generated code: a class named \codefont{BcryptPasswordHashFunctionTest} defines a test function \codefont{testCheckFunction()}, to call the vulnerable API with  
appropriate formats of the seed inputs of exemplar test. The generated test is very similar to the exemplar test, but it shows 
PoV for client code.


\begin{figure}
 \begin{mybox}
    {\textbf{\textit{Prompt Example}}}
Write a JUnit test for function: \textit{[check]} in "CLIENT CODE``. The test should verify whether the function \textit{[check]} is affected by the vulnerability: \textit{[CVE-2020-28052]}, or affected by vulnerable APIs: \textit{[OpenBSDBZCrypt.checkPassword]}. Mimic the test function "TEST FUNCTION:`` \textit{[performTest]} I provide to generate the input in your test. For the provided code "CLIENT CODE``, mock the classes and functions that are not Java SE APIs and have no definition provided.
\begin{verbatim}
TEST FUNCTION:
``` Java
// Here we omit details of BC security test 
// (Listing 1) for brevity.
@Test
public void performTest() throws Exception { ... }
```

CLIENT CODE: 
``` Java
// Here we omit details of the class, which 
// defines the method check (...) in Listing 2
public class BcryptPasswordHashFunction 
    implements PasswordHashFunction { ... }
```
\end{verbatim}
\end{mybox}
\vspace{-1em}
\caption{A prompt derived from the default template}\label{fig:prompt-example}
\vspace{-1.5em}
\end{figure}

\begin{figure}[b]
\caption{A brief and commented version of the security test successfully generated by ChatGPT.}
\label{lst:generate}\vspace{-2em}
\begin{lstlisting}[style=javastyle]
... // We omit less important details for brevity
public class BcryptPasswordHashFunctionTest {
  ...   
  @Test
  public void testCheckFunction() {
    int costFactor = 4;
    for (int i = 0; i < 1000; i++) {
      random.nextBytes(salt);
      final String tokenString = OpenBSDBCrypt.generate("test-token".toCharArray(), salt, costFactor);
      assertTrue(bcryptPasswordHashFunction.check( tokenString, "test-token"));
      /* The App should fail the following assertion, when it depends on a vulnerable BC version that messes up correct with incorrect passwords.*/      
      assertFalse(bcryptPasswordHashFunction.check( tokenString, "wrong-token"));
} } }
\end{lstlisting}
\vspace{-.5em}
\end{figure}

\subsubsection{The Design of Alternative Prompt Templates}\label{sec:variants}
In addition to the default prompt template, we also defined five variant templates by removing a single element from (iii)--(vii) each time. In this way, we can explore how different information elements contribute to \tool's effectiveness. 
As shown in Table~\ref{tab:templates}, to facilitate presentation, we use $P$ to  refer to the default template, and 
use $P_1$--$P_5$ to refer to the five variants. 
For instance, $P_5$ leaves out (vii), but takes in (iii)--(vi). 
We then generated \totalClient prompts using each variant template.

\subsection{Phase III: Result Validation}\label{sec:validation}
After sending all prompts to \tool, we gathered and recorded \tool's outputs. We integrated the generated tests into Apps using the following rules: if a generated test class $C'$ has a unique name and defines one or more test functions for method-to-test $M$, we place the class into the test folder.  Otherwise, if $C'$ has the same name as an existing class, we put the non-conflicting content into the existing class; if $C'$ defines test functions for a non-public method $M$, we place the tests into any existing (test) class where $M$ is accessible. 

\begin{table}
\scriptsize
\centering
\caption{The six prompt templates we explored}\label{tab:templates}\vspace{-.5em}
\begin{tabular}{l |l ?l |l}
\toprule 
\textbf{Id} &\textbf{Prompt Template} &\textbf{Id} &\textbf{Prompt Template}\\ \toprule
$P$ &Default (including iii-vii) & $P_3$ &Without C (v)\\ \hline
$P_1$ &Without vulnerable APIs (iii) &$P_4$ &Without the exemplar test (vi)\\ \hline
$P_2$ &Without $M$ (iv) & $P_5$&Without the vulnerability ID (vii)\\  \bottomrule
\end{tabular}\vspace{-2em}
\end{table}

For each generated test, we compiled App with the test to check for compilation errors; if some compilation errors were obvious and easy to fix (e.g., missing/wrong package names or missing library dependencies), we fixed those errors manually to explore whether \tool could synthesize the most important logic of security tests. After one or multiple iterations of the compilation-and-fixing procedure, if a test compiled successfully, we further executed App with that test to observe runtime behaviors. If any exception or runtime error was thrown, we studied the exception/error message, inspected the intermediate program status via step-by-step debugging, and discussed the relevance with PoV among authors until reaching a consensus.

\section{Experiments and Results}

We empirically investigated three research questions (RQs): 
\begin{itemize}
\item[\textbf{RQ1:}] \emph{How effectively does \tool generate security tests?} We explored the strengths and weaknesses of \tool in security test generation. 
\item[\textbf{RQ2:}] \emph{How does ChatGPT's security test generation performance differ given various types of prompts?} 
Among the information elements (iii)--(vii), we explored which one is more crucial. 
\item[\textbf{RQ3:}] \emph{How does \tool compare with existing tools of security test generation?} 
We wanted to learn how well LLM-based security test generation compares with existing tools based on program analysis.
\item[\textbf{RQ4:}] \emph{How does \tool work with few-shot prompting?} We explored whether \tool works better when one or more examples of test-generation tasks are provided. 

\item[\textbf{RQ5:}] \emph{How effectively do different LLMs in generate security test?} We investigated whether our findings generalize across different LLMs by evaluating the performance of five state-of-the-art models on the security test generation tasks. 

\end{itemize} 
This section first introduces the metrics we defined to assess tools of security test generation (Section~\ref{sec:metrics}). It then explains our experiments and results for RQs. 

\subsection{Metrics}\label{sec:metrics}

There are three metrics used in our experiments:

\textbf{Tool Applicability (A)} counts for how many $\langle App, Lib\rangle$ pairs,  
a tool generates a security test. 

\textbf{Test Compilability (C)} counts the number of generated tests that are compilable, with minor fixes applied. 

\textbf{PoV Demonstration (D)} counts the number of compilable tests that successfully demonstrate PoV for
the known vulnerabilities as expected.

\begin{table*}
\scriptsize
\centering
\caption{Security tests generated by \tool (Tool Applicability (A): \totalClient, Test Compilability (C): \totalCompTest, PoV Demonstration (D) \completeECount)}:  \label{tab:rq1}\vspace{-.5em}
\begin{tabular}{r| l| R{3em}|R{1em}| R{1em}| R{1em}?
l| l| R{3em}|R{1em}| R{1em}| R{1em}}
\toprule
\textbf{Idx} &\textbf{Vulnerability Entry ID} &\textbf{\# of Clients}&\textbf{A} &\textbf{C}  &\textbf{D} 
&\textbf{Idx} &\textbf{Vulnerability Entry ID} &\textbf{\# of Clients}&\textbf{A} &\textbf{C}  &\textbf{D}\\ \toprule
1 & CODEC-134 & 3 & 3 & 2  & 2 
 &15 &CVE-2020-13973 &1 &1 &0& 0\\ \hline
2 & CVE-2017-7525 & 2&2 & 2  & 2
 & 16 &CVE-2020-26217 &3 &3&3 &2 \\ \hline
3 & CVE-2017-7957 & 2 &2 & 1&1
 &17 & CVE-2020-28052 & 2 &2& 2  & 1 \\ \hline
4 & CVE-2018-1000632 & 2 & 2 &1  &1 
 & 18 & CVE-2020-28491 &1 &1 &1  &0\\ \hline
5 & CVE-2018-1000873  & 3 &3 & 2  & 0
 & 19 &CVE-2020-5408 &2 &2& 2& 2*\\ \hline
6 & CVE-2018-1002200 & 2 & 2 &1 & 1 &
 20 & CVE-2021-23899 &1 &1&1&0\\ \hline
7 &CVE-2018-1002201 &1& 1 & 1& 1 &
 21 &CVE-2021-27568 & 2&2 & 2& 2 \\ \hline
8 & CVE-2018-11761 & 1&1 & 0 & 0 
 & 22 &CVE-2021-29425 &3 &3 & 2 & 2\\ \hline
9 &CVE-2018-12418 &1&1&0  & 0
 & 23 &CVE-2021-30468 & 1& 1& 1    &0\\ \hline
10& CVE-2018-1274 &1 &1&0 & 0
 & 24 & CVE-2022-25845 & 2 &2 & 2& 2\\ \hline
11 &CVE-2018-19859 &1 &1&1&0 
 & 25 & CVE-2022-45688 & 3 &3 & 3 & 1\\ \hline 
12 
&CVE-2019-10093 & 1 &1 &1  &0 
 & 26 &CVE-2023-34454 & 1&1&1  &1\\ \hline
13
&CVE-2019-12402 &1 &1&0  & 0
 & 27 & HTTPCLIENT-1803 &1 & 1 &1 &0 \\ \hline
14
&CVE-2020-13956 &1 &1 &1 &1 
 & 28 & TwelveMonkeys-595 &2 &2& 2& 0\\ \hline

\multicolumn{5}{c}{}&& 29 & Zip4j-263 &2 & 2 &2& 2
\\ \bottomrule 
\multicolumn{4}{l}{*indicates the number of clients with injected vulnerable dependencies}\\
\end{tabular}
\vspace{-1.em}
\end{table*}

\subsection{\tool's Effectiveness in Security Test Generation (RQ1)}\label{sec:rq1}
We created \totalClient prompts using the default prompt template $P$, and sent them to \tool \tse{, which uses \gpt as an underlying model in our experiment}. 

As shown in Table~\ref{tab:rq1}, \tool generated tests for all prompts. Of these, \runnable tests are compilable and runnable as is. In contrast, \minorfix tests become compilable and runnable after we manually apply minor fixes, such as adding missing dependencies, handling exceptions, or correcting hardcoded file paths.  
\completeECount \tse{(including five tests with minor fixes)} of these \totalCompTest tests effectively mimic the behaviors of given library tests, and successfully demonstrate PoV by throwing relevant errors or runtime exceptions.

\subsubsection{Uncompilable Tests.} Among the \totalClient generated tests, \notCompTest tests do not compile and cannot be easily fixed via minor changes. Specifically, 
two of the tests violate Java access rules, such as directly accessing private members outside of declaring classes.
Five tests use undefined program entities (e.g., methods and classes).
Another four tests 
call methods inappropriately, by missing some parameters or using the wrong type of parameters. Our observations imply that \tool does not guarantee code compilation, even though the majority of tests it generated (\totalCompTest/\totalClient) are easy to compile. 

\subsubsection{Less Effective Tests.} 14 generated tests do not trigger vulnerabilities as expected. They either throw exceptions/errors other than the expected ones, or throw no exception/error at all. 
Three reasons can explain such ineffectiveness. 
First, Mockito~\cite{mockito}---a mocking framework---was used by \tool to mock unknown variables, methods, or classes when generating tests. Unfortunately, the framework could not mock everything (e.g., final classes), and sometimes led to MockitoExceptions.  
Second, in generated tests, the parameters passed to $M$ are not always well prepared. They may be malformed, include \codefont{null}-values in critical fields, or fail to contain the essential values to trigger vulnerabilities. 
Third, our generated tests uncovered two unexpected bugs. Specifically, one test exposed a concurrency issue in \codefont{NCI-Agency/anet's} \codefont{sanitizeJson()}, leading to a potential denial-of-service vulnerability in applications. This issue has been reported and assigned CVE-2023-31441. The other bug initially presented as a \codefont{NullPointerException}. However, upon inspecting the exception chain, we discovered that the test revealed a Server-Side Request Forgery (SSRF) vulnerability within the AssetVersionTransformer component of the \codefont{aem-caching} library (version 0.0.3-SNAPSHOT).

\subsubsection{Effective Tests.} 24 generated tests can trigger vulnerabilities as expected, including two cases involving projects with injected vulnerable dependencies. For these tests, \tool successfully extracted vulnerability-triggering inputs from the exemplar tests, reused those inputs to call method $M$ somehow, and presented relevant abnormal program behaviors (e.g., infinite loop). As most vulnerabilities were fixed in the latest versions of the subject projects, we only filed reports for four of the newly revealed vulnerabilities in Apps. Two of them were approved: CVE-2023-37760 and CVE-2023-43151.

Based on our experience, when $M$ calls vulnerable API(s) directly and defines a parameter list the same as the called API(s), \tool was more likely to succeed in producing security tests;  15 of the 24 tasks included such $M$ (i.e., methods under test).
This is understandable as \tool does not reason about program logic. When it tries to generate a test similar to a given test, any commonality among the App context, vulnerable API, and exemplar test can facilitate knowledge reuse and test mimicry. 

Meanwhile, when $M$ calls vulnerable API(s) indirectly or defines a different parameter list from the called API(s), \tool was less capable because less commonality is shared between App and Lib tests.
Interestingly, among the 24 tasks well handled by \tool, 9 tasks involve $M$ with a different parameter list and 1 task defines $M$ to indirectly call the vulnerable API.
It implies three things. 
First, 
\tool is promising to effectively generate tests, even though the vulnerability propagation path from Lib to App is more complex or less obvious.
Second, \tool's effectiveness decreases as the test-generation tasks become more challenging.
For the three categories of functions (i.e., C1--C3) in our dataset, \tool's success rates of demonstrating PoV
are 75\% (15/20), 45\% (9/20), and 11\% (1/9). Namely, 
\tool is more effective in generating tests when $M$ shares more commonality with vulnerable APIs, but less effective when there is less commonality. 
Third, our experiment results are generalizable, as our dataset is not limited to trivial cases where functions-under-test and vulnerable APIs share both parameter lists and program context.
Namely, the more commonality between $M$ and vulnerable APIs (category C1), the more likely \tool can generate security tests successfully.

\vspace{0.3em}
\noindent\begin{tabular}{|p{8.5cm}|} 
	\hline
	\textbf{Finding 1:} \emph{\tool is promising in generating security tests for known library vulnerabilities. Given \totalClient test generation tasks, it produced \totalClient tests, \totalCompTest of which are easy to compile and \completeECount tests successfully demonstrated PoV.
	}
	\\
	\hline
\end{tabular}
\vspace{0.2em} 

\begin{table*}
\scriptsize
\centering
\caption{The comparison of tests generated in different ways}\label{tab:rq2}
\vspace{-.5em}
\begin{tabular}{l| l| r| R{.5cm}| R{1.cm} R{1.cm} R{1.2cm} | R{.5cm}| R{1.cm} R{1.cm} R{1.3cm}}
\toprule
\multirow{3}{*}{\textbf{Id}} &\multirow{3}{*}{\textbf{Prompt Template}} &{\textbf{Tool}} &\multicolumn{4}{c|}{\textbf{Test Compilable?}}
&\multicolumn{4}{c}{\textbf{PoV Demonstrated?}} \\ \cline{4-11}
& &\textbf{Applicability}  &{\textbf{Yes}} & \multicolumn{3}{c|}{\textbf{No}}
&{\textbf{Yes}} & \multicolumn{3}{c}{\textbf{No}}
\\ \cline{5-7}\cline{9-11}
& & \textbf{(A)}& \textbf{(C)}&
\textbf{Access Rule Violation} & \textbf{Incorrect Method Calls} &\textbf{Unknown Entity Usage} &
\textbf{(D)} & \textbf{No error/ exception} &\textbf{Mockito exception} &\textbf{Other exceptions/ errors}\\ \toprule
$P$ &Default (all elements) & \totalClient & \totalCompTest& \ying{2} &4&5 & 24 & \ying{3} &2 & \ying{9}\\ \hline
$P_1$&Without vulnerable APIs (iii) & \totalClient & \ying{34}&3 &2& \ying{10} & 15&6&8& \ying{5}\\ \hline

$P_2$ & Without $M$ (iv) & \totalClient & \ying{\noM} & \ying{4} & 1 & \ying{5} & 14 & \ying{3} & 8 & \ying{14}\\ \hline
$P_3$ &Without C (v) & \totalClient & \ying{\noC} &3 & 1 &\ying{30} & 1 & 7 & \ying{4}& \ying{3} \\ \hline
$P_4$&Without the exemplar test (vi) & \totalClient &  \ying{30} & \ying{4} &2 & \ying{13} &0 &11&11& \ying{8}\\ \hline
$P_5$ &Without the vulnerability ID (vii) & \totalClient & \ying{ \noCVE} & \ying{6} & 0 & 7 & 14 & \ying{3} &5&\ying{14}\\ \bottomrule
\end{tabular}
\vspace{-2em}
\end{table*}  

For the 29 vulnerabilities we investigated, the 29 exemplar tests have 5 types of test oracles: thrown exceptions, thrown runtime errors, timeouts, infinite loops, and expected return-values of function calls. Among those exemplar tests, when vulnerable functions are called, 2 tests throw expected exceptions; 10 tests do not throw exceptions as expected; 4 tests throw runtime errors; 1 test does not throw an error as expected; 3 tests encounter time out; 4 tests run into infinite loops; 2 tests do not output expected values; 3 tests produce the expected problematic outputs.

Among the 24 successfully generated tests, 22 of them follow similar vulnerability exploitation logic as the given exemplar tests. In particular, when vulnerable functions are called, 3 of the generated tests throw expected exceptions; 13 tests do not throw exceptions as expected; 3 tests throw runtime errors, 1 test does not throw an error as expected; 1 test encounters time out; 2 tests do not output expected values; 1 test produces the expected problematic output. None of the generated tests run into infinite loops, which implies that it may be harder for ChatGPT to generate tests to demonstrate such abnormal program behaviors. 

\vspace{0.3em}
\noindent\begin{tabular}{|p{8.5cm}|} 
	\hline
	\textbf{Finding 2:} \emph{\tool effectively mimics human-crafted tests;  
    22 of the 24 tests it successfully generated have matching logic with the given exemplar tests.}
	\\
	\hline
\end{tabular}
\vspace{0.2em} 

To check the consistency of \tool's outputs, 
we conducted an experiment by sending 25 test-generation tasks 5 times to ChatGPT, using 125 distinct conversation sessions. 
We randomly sampled those tasks, without bias towards any data. 
\tool produced consistent results 113 of 125 times (90\%). Our results imply that \tool often produces very similar results, given the same prompt multiple times.

\subsection{Impact of Information Elements on \tool's Outputs (RQ2)
}\label{sec:rq2}

We used template variants $P_1$--$P_5$ to generate prompts for \totalClient $\langle App, Lib\rangle$ pairs, and sent all prompts to \tool for results.  

\subsubsection{\tool's Applicability Given Divergent Types of Prompts}
As shown in Table~\ref{tab:rq2}, no matter what information item in the default template was removed, the resulting prompts always guided \tool to produce tests. It means that \tool has great applicability: it is always applicable no matter how the prompts were formulated. 

\subsubsection{\tool's Test Compilability Given Divergent Types of Prompts}
As shown in Table~\ref{tab:rq2}, when $P_2$ was used and M was not specified, \tool generated more compilable tests than it did for the default template $P$ (\noM vs.~\totalCompTest).
However, when $P_3$ and $P_4$ were used, a lot fewer generated tests are compilable, i.e., \ying{\noC and \noExaTest}. 
One possible reason is that 
 both $P_3$ and $P_4$ removed the code context, while the other templates removed almost no code context. 
 As a generative AI tool, \tool predicts the next word(s) given a data sequence, by using (1) an encoder to process the input sequence and (2) a decoder to generate the output~\cite{inside-chatgpt}. 
It tended to generate more compilable tests when more relevant program context was provided.

Among the 6 prompt templates, $P_3$ caused \tool to create the most uncompilable tests---\tse{34}; \ying{30} of these tests fail compilation due to their usage of unknown entities. 
This may be because $P_3$ does not specify the Java class C holding the function-to-test;
 \tool could not identify many valid or usable entities available in the software projects, so it usually refers to some nonexistent entities in the produced tests. In contrast, $P_2$ caused \tool to create the fewest uncompilable tests---10, only 5 of which fail compilation due to their usage of unknown entities. This comparison implies that \tool could produce more compilable tests when (1) more program context is provided in prompts, and (2) there is no constraint on what method to test. 

No matter what template we used, \tool always produced uncompilable tests for some prompts. This observation indicates that \tool does not strictly follow Java rules on syntax or semantics. 
As a generative AI model, 
it was trained to predict the next words or phrases to follow a given sequence. 
Thus, the generated code may violate access rules, call methods with inappropriate parameter lists, or use unknown program entities. This observation implies the necessity of applying sanity checks to \tool-generated code and fixing any revealed bugs, to ensure the program quality.

\vspace{0.3em}
\noindent\begin{tabular}{|p{8.5cm}|}
	\hline
	\textbf{Finding 3:} \emph{Among the five template variants we explored, \tse{$P_3$ (Without the client project class C)} and \tse{$P_4$ (Without the exemplar test)} caused \tool to work considerably worse in producing compilable tests. \tool tended to produce more compilable tests, given more contextual code and fewer constraints relevant to the test-generation tasks.	}
	\\
	\hline
\end{tabular}
\vspace{0.2em} 

\subsubsection{\tool's PoV Demonstration Given Divergent Types of Prompts}
Removing any element from $P$ worsened \tool's effectiveness. Among the variants, $P_4$ caused \tool to work worst, producing zero successful PoV demonstration. This phenomenon implies that exemplar tests offer (1) important program structures for potential security tests, and (2) essential hints on vulnerability-triggering inputs. Without Lib tests, \tool generated 11 tests throwing no error/exception, 11 tests wrongly mocking program entities, and \ying{8} tests triggering irrelevant errors/exceptions.
Although slightly better than $P_4$, 
$P_3$ also worsened \tool significantly and only one security test was produced successfully.
This may be because the removed Java class C holds lots of context, whose absence caused \tool to create tests in a context-agnostic way, making the created tests irrelevant or invalid.

$P_1$, $P_2$, and $P_5$ had very similar effects on \tool, as the tool produced 15, 14, and 14 security tests given the prompts derived from each of them. All these numbers are much lower than the number reported for the default template $P$: 24.
This implies that the elements removed by individual templates (iii, iv, vii) provide valuable signals to \tool, to help it identify and focus on the vulnerable APIs, function-to-test, and specialized vulnerability.
While (v) and (vi) provide as much relevant code as possible for \tool to refer to, 
the other elements (iii, iv, vii) guide \tool to pay special attention to the most important content in the relevant code.

\vspace{.3em}
\noindent\begin{tabular}{|p{8.5cm}|}
	\hline
	\textbf{Finding 4:} \emph{Among the five information elements covered by the default prompt template $P$, all elements played an important role to help \tool effectively generate security tests.
    In particular, (v) and (vi) were more important than (iii), (iv), and (vii). 
	}
	\\
	\hline
\end{tabular}
\vspace{.2em} 
\vspace{-1em}
\subsection{Tool Comparison (RQ3)}\label{sec:rq3} 

\begin{table}
\scriptsize
\caption{The input information required by the default setting of different tools (Yes: \cmark, No: \xmark)}\label{tab:tool-input}
\centering
\begin{tabular}{L{2.5cm}|R{1.cm}|R{1.5cm}|R{1.5cm}}
\toprule
\textbf{Information} & \textbf{SIEGE} &\textbf{TRANSFER} &\textbf{\tool} \\ \toprule
Vulnerable API & \cmark &\cmark&\cmark\\ \hline
$M$ &\xmark &\cmark&\cmark\\ \hline
$C$&\xmark&\cmark&\cmark\\ \hline
Exemplar test &\xmark &\cmark&\cmark\\ \hline
Vulnerability ID&\xmark &\xmark&\cmark\\ \hline
All .class files of App &\cmark &\cmark &\xmark\\ \hline
JAR file of Lib &\cmark&\cmark&\xmark\\ \hline
Vulnerable line number in Lib &\cmark &\xmark &\xmark\\ 
\bottomrule
\end{tabular}
\end{table}

Two tools were recently proposed to automatically generate security tests: SIEGE~\cite{iannone2021toward} and TRANSFER~\cite{kang2022test}. 
SIEGE adopts a genetic algorithm (GA). 
 For any $\langle App, Lib \rangle$ pair, 
 SIEGE must be executed under the project folder of App, which includes all compiled .class files of App and JAR files of library dependencies (including Lib). As shown in Table~\ref{tab:tool-input}, SIEGE also requires users to specify the search target (i.e., the coverage goal for tests-to-generate), including the vulnerable API and vulnerable line number in Lib. 
  SIEGE reuses EvoSuite~\cite{fraser2011evosuite}---the popularly used test generation tool---to generate tests, select tests based on their closeness to the specified target, and evolve those tests with some randomness to get better tests.
  SIEGE stops when the time budget is used up or some tests perfectly match the target.

Similar to SIEGE, TRANSFER also generates security tests using GA. 
However, as shown in Table~\ref{tab:tool-input}, TRANSFER requires users to provide  slightly different inputs: instead of including the vulnerable line number, users should designate $C$, $M$, and an exemplar test from Lib. 
It conducts program static analysis
to create both call graphs and control-flow graphs for App, and uses dynamic instrumentation 
to assess test coverage.
It executes and dynamically instruments the exemplar test, 
to identify program states relevant to the vulnerability, and to extract conditions that must be satisfied by any generated security test. Finally, TRANSFER  adopts the extracted information to guide EvoSuite and derive security tests. Due to the usage of exemplar test and advanced program analysis techniques, TRANSFER manifested better effectiveness than SIEGE~\cite{kang2022test}. 

\subsubsection{Experiment}
We prepared the inputs required by SIEGE and TRANSFER for \totalClient $\langle App, Lib \rangle$ pairs, and executed both tools. 
Note that the two tools have full access to the .class files of each App and the JAR file of each Lib, while \tool can only access the partial code described in prompts.
{For accurate comparison, we properly prepared the inputs for individual tools. 
We copy-and-pasted each tool-generated test to appropriate places. We leveraged the build process to reveal compilation errors. We also applied minor fixes to obvious and simple compilation errors.
If compilation succeeded, we executed App with the generated test to observe runtime behaviors. If any exception or runtime error was thrown, we studied the exception/error message, inspected intermediate program states via step-by-step debugging, and discussed the relevance among authors until reaching a consensus.

\begin{table}
\centering
\scriptsize
\caption{The comparison between \tool and state-of-the-art tools on our dataset}\label{tab:rq3}
\vspace{-.5em}
\begin{tabular}{ L{2cm}| R{1.5cm} r R{1.7cm}}
\toprule
 &\textbf{Tool Applicability} &\textbf{Test Compilability} &\textbf{PoV} \\ 
 &\textbf{(A)} &\textbf{(C)} &\textbf{Demonstration (D)} \\ \toprule
\tool & \totalClient & \totalCompTest (\runnable + \minorfix$^*$)& \completeECount (19 + 5$^*$)\\ \hline
\tool w/o vulnerability ID (i.e., $P_5$) &49&36 (18 + 18*) &14 (7 + 7*)\\ \hline
TRANSFER &16  & 13 (9 + 4$^*$) & 4 (3 + 1$^*$) \\ \hline
SIEGE& 1 & 1 & 0\\ \bottomrule
\multicolumn{4}{l}{* marks the number of tests that compile after we applied minor fixes}\\ \bottomrule
\end{tabular}
\vspace{-2em}
\end{table}

\subsubsection{Results} As shown in Table~\ref{tab:rq3}, \tool outperformed current tools considerably by having much better tool applicability, test compilability, and PoV demonstration. It generated tests for all \totalClient $\langle App, Lib\rangle$ pairs, while TRANSFER only generated test functions or code snippets for 16 pairs. SIEGE worked much worse, creating a test for only one pair.

In terms of compilability, 13 out of the 16 tests generated by TRANSFER are compilable, including 4 tests with minor fixes applied;  3 of the 16 tests fail compilation due to their usage of unknown entities. For those 16 tasks, \tool generated 15 compilable tests, including 3 tests fixed with minor edits. 
The only test output by SIEGE compiles successfully. For that same task, \tool also generated a compilable test. These phenomena imply that \tool outperforms existing tools by generating more compilable or easy-to-compile tests. There is no task showing either existing tool to outperform \tool.

In terms of PoV demonstration, only four of the tests output by TRANSFER trigger vulnerabilities; For the four Apps handled well by TRANSFER, \tool also generates tests that trigger the same vulnerabilities, indicating that it produces a superset of the effective PoVs found by TRANSFER.
Among the remaining nine compilable tests produced by TRANSFER,  six tests execute smoothly, without any error or exception; three tests trigger irrelevant exceptions. The only test by SIEGE fails to trigger any vulnerability, because it throws an irrelevant exception. These observations indicate that \tool not only outperforms TRANSFER but also discovered more diverse and effective vulnerability-triggering PoV test

When \tool and TRANSFER perform differently, one may wonder 
whether (1) the extra input of vulnerability ID that \tool takes or (2) the distinct tools' working mechanisms explain the observed differences. To characterize the contributions of both factors between \tool and TRANSFER, in Table~\ref{tab:rq3}, we also included a variant approach of the default \tool usage by using template $P_5$. As shown in the table, when \tool took in part of its inputs that are also required by TRANSFER, it achieved better tool applicability (49 vs.~16), test compatibility (36 vs.~13), and PoV demonstration (14 vs.~4) than TRANSFER. Meanwhile, the variant approach worked less effectively than our default \tool usage. These results imply that between \tool and TRANSFER, both (1) the vulnerability ID that \tool takes and (2) the distinct tools' working mechanisms contribute to the observed differences, although (2) plays a more important role.

We further inspected the cases where TRANSFER worked worse than \tool, and summarized three major limitations of the tool design.
First, TRANSFER adopts EvoSuite,} to generate tests and execute the method-to-test $M$. However, EvoSuite is not quite effective; some or even most of the tests generated by EvoSuite do not execute M at all. Second, TRANSFER has difficulty synthesizing or mocking complex input parameters for $M$. For instance, it could not  synthesize a parameter of type \codefont{net.sourceforge.pmd.RuleSet}, but \tool mocked such an object via the Mockito framework. 

Third, TRANSFER has difficulty incorporating the knowledge embedded in exemplar tests into test generation. For instance, CODEC-134 is related to two vulnerable APIs: \codefont{Base64.decode(...)} and \codefont{Base64.decodeBase64(...)}. 
We provided both TRANSFER and \tool inputs relevant to that vulnerability, including the Lib test to show PoV of one API \codefont{Base64.decode(...)}, 
and a client Java class to call the other API \codefont{Base64.decodeBase64(...)}. 
We used both tools to generate a security test for the client code. 
Unfortunately, TRANSFER could not reuse any domain knowledge from the Lib test, but \tool successfully achieved that.

SIEGE worked much worse than \tool. The major reason is that SIEGE has very limited applicability, probably due to implementation issues. Among the 49 $\langle App, Lib\rangle$ pairs, SIEGE was only applicable to a single pair.

\vspace{0.3em}
\noindent\begin{tabular}{|p{8.5cm}|}
	\hline
	\textbf{Finding 5:} \emph{\tool outperformed both TRANSFER and SIEGE on our dataset. No security test was successfully generated by TRANSFER or SIEGE, but not by \tool.
 \tool has great potentials in generating security tests.
	}
	\\
	\hline
\end{tabular}
\vspace{0.3em}

\subsubsection{Additional Experiments and Results}
For fair comparison, we also tried to apply tools to the datasets mentioned by papers of SIEGE and TRANSFER~\cite{iannone2021toward,kang2022test}. 
Unfortunately, as we mentioned before,
the open-sourced dataset of TRANSFER~\cite{transfer-dataset} {has 21 apps that are not quite usable in our evaluation.
Seven of the Apps lack essential details for successful build or execution; six Apps 
depend on secure instead of vulnerable versions of Libs; five Apps have no vulnerable API calls or Lib test; two Apps wrap vulnerable API calls with security sanity checks to eliminate potential exploits; one App calls the vulnerable API in the test file. To fully leverage TRANSFER's dataset in our tool-comparison experiment, we manually replaced the security dependencies with vulnerable ones in six Apps, and still used the single App that calls vulnerable API in the test file.
}

\begin{table}
\centering
\scriptsize
\caption{The comparison between \tool and TRANSFER on the TRANSFER's dataset that we restored}\label{tab:rq3-2}
\vspace{-.5em}
\begin{tabular}{ L{2cm}| R{1.5cm} r R{1.7cm}}
\toprule
 &\textbf{Tool Applicability} &\textbf{Test Compilability} &\textbf{PoV} \\ 
 &\textbf{(A)} &\textbf{(C)} &\textbf{Demonstration (D)} \\ \toprule
 
\tool & 28 &20 (9 + 11*)& 16 (9 + 7*)\\ \hline
ChatGPT w/o vulnerability ID (i.e., $P_5$) & 28 & 24 (13+11*) &14 (10+4*)\\ \hline
TRANSFER &12  & 9 (3 + 6*)& 5 (3 + 2*) \\ \bottomrule
\end{tabular}
\vspace{-2em}
\end{table}
 
{In this way, we
experimented with 28 of the 42 $\langle App, Lib\rangle$ pairs in TRANSFER's dataset.} 
As shown in Table~\ref{tab:rq3-2}, by applying \tool and TRANSFER to that dataset, we found \tool  outperform TRANSFER in all metrics. We also applied \tool's variant approach based on $P_5$, as this variant takes only inputs that the default approach commonly shares with TRANSFER. This variant also outperformed TRANSFER in all metrics, indicating the stronger power of \tool in generating security tests.

\vspace{0.3em}
\noindent\begin{tabular}{|p{8.5cm}|}
	\hline
	\textbf{Finding 6:} \emph{\tool outperformed TRANSFER on the TRANSFER's dataset that we restored, no matter whether \tool takes in solely the inputs shared with TRANSFER or those together with the unique input vulnerability ID. 
	}
	\\
	\hline
\end{tabular}

\begin{table}[h]
\centering
\scriptsize
\caption{The comparison between \tool and SIEGE on SIEGE's dataset}\label{tab:rq3-3}
\vspace{-.5em}
\begin{tabular}{ L{2cm}| R{1.5cm} r R{1.7cm}}
\toprule
 &\textbf{Tool Applicability} &\textbf{Test Compilability} &\textbf{PoV} \\ 
 &\textbf{(A)} &\textbf{(C)} &\textbf{Demonstration (D)} \\ \toprule
\tool w/o exemplar test (i.e., $P_4$) & 11 & 10 & 0\\ \hline
\tool taking in only vulnerable API and $C$ & 11 &7&0 \\ \hline
SIEGE & 9 & 9&0 \\ \bottomrule
\end{tabular}
\vspace{-1.em}
\end{table}

Although SIEGE's dataset is publicly available, the dataset includes no Lib test for \tool to refer to. Therefore, when applying \tool to SIEGE's dataset, by default, our prompts do not include any Lib test. 
Furthermore, we also explored a variant usage of \tool by specifying only the vulnerable API and client code $C$, so that this variant takes in no more input than SIEGE.

As shown in Table~\ref{tab:rq3-3}, SIEGE generated tests for 9 of the 11 tasks; the tests are compilable, but no test shows PoV. In comparison, \tool-without-exemplar-test generated 11 tests; 10 of the tests are compilable, and 1 test is uncompilable due to wrong type casting. None of the tests show PoV. This is as expected. 
As mentioned in Section~\ref{sec:rq2}, 
when no exemplar test is offered, \tool cannot effectively generate security tests.
\tool's variant usage that takes no more input than SIEGE also generated 11 tests; 7 of the tests are compiable and 0 test shows PoV. The result implies that compared with SIEGE, \tool always has a better tool applicability, although the test compilability and PoV demonstration of \tool-generated tests are not necessarily better.

\vspace{0.3em}
\noindent\begin{tabular}{|p{8.5cm}|}
	\hline
	\textbf{Finding 7:} \emph{\tool is always more applicable than SIEGE. However, when no exemplar test is provided, \tool does not achieve better PoV demonstration.
	}
	\\
	\hline
\end{tabular}
\subsection{Comparison between Zero-shot Prompting and Few-shot Prompting (RQ4)}
\label{sec:rq4}

\begin{figure}
    \centering
    \includegraphics[width=\linewidth]{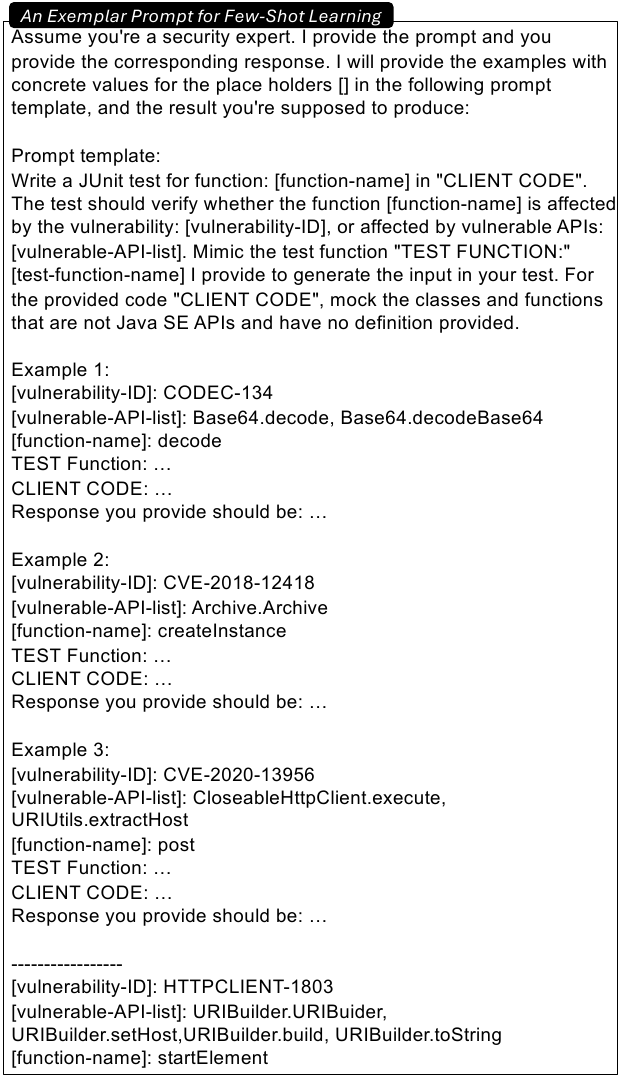}
    \vspace{-1.5em}
    \caption{A three-shot prompt that provides examples for three test-generation tasks}
    \label{fig:few-shot}
\end{figure}

In all experiments we have explained so far, we adopted the zero-shot prompting technique. Namely, we did not provide any exemplar question-answer pair to \tool, to enable in-context learning where demonstrations are provided in prompts to steer the model for better performance. 
One hypothesis we had is that \tool might work better given few-shot prompts. To validate that hypothesis, we conducted an experiment by sending \tool question-answer pair(s) in each prompt. As shown in Fig.~\ref{fig:few-shot}, a three-shot prompt first describes the default template $P$ we used for zero-shot learning. It then provides three pairs of (1) input parameters to customize $P$, and (2) exemplar output we manually crafted in response to those customized test-generation tasks. Finally, it offers a fourth set of input parameters to customize $P$, making \tool generate a test accordingly.

To improve the representativeness of our evaluation, we experimented with two types of few-shot prompts: one-shot and three-shot prompts. To avoid the randomness contributed by diverse question-answer pairs, for each type of prompts, we intentionally provided the same set of question-answer pair(s). Namely, among the 49 $\langle App, Lib\rangle$ pairs in our dataset, for 3-shot prompts, 
we randomly picked 3 pairs for example construction, and used the remaining 46 pairs for evaluation.
For one-shot prompts, we used one of the three pairs for example  construction, and reused the evaluation set for three-shot prompts mentioned above. 
By splitting the dataset in this way, we ensure that (1) there is no overlap between the exemplar question-answer pairs and evaluation questions, and (2) different prompting techniques are evaluated with exactly the same data portion.

\begin{table}
\scriptsize
    \caption{The comparison between zero-shot and few-shot prompting of \tool on the dataset of 46 $\langle App, Lib\rangle$ pairs}
    \label{tab:few-shot}
    \centering
\begin{tabular}{ L{2cm}| R{1.5cm} r R{1.7cm}}
\toprule
 &\textbf{Tool Applicability} &\textbf{Test Compilability} &\textbf{{PoV}} \\ 
 &\textbf{(A)} &\textbf{(C)} &\textbf{{Demonstration (D)}} \\ \toprule
\tool with zero-shot learning (default) &
\totalGPTClient & 36 (24 + 12*)& 22 (17 + 5*)\\ \hline
\tool with one-shot learning &\totalGPTClient & 32 (22 + 10*)& 14 (11 + 3*) \\ \hline
\tool with three-shot learning &
\totalGPTClient & 37 (30 + 7*) & 24 (23 + 1*)\\ 
 \bottomrule
\end{tabular}
\vspace{-2em} 
\end{table}

Table~\ref{tab:few-shot} compares the results of few-shot prompts against those of our default prompts. As shown in the table, with one-shot prompts, \tool has worse test compilability (32 vs.~36), and worse PoV demonstration (14 vs.~22). However, with three-shot prompts, \tool has slightly better test compilability (37 vs.~36) and better PoV demonstration (24 vs.~22). Surprisingly, including question-answer pairs into prompts does not guarantee improvements, but may worsen the results. Two possible reasons can explain the observed phenomena. First, when one-shot prompts are provided, the only question-answer pair mentioned is insufficient to steer \tool for better performance, but may confuse the model to output less optimal results. Second, when three-shot prompts are offered, the multiple question-answer pairs mentioned present more demonstrations to better enable in-context learning. 
\vspace{0.3em}
\noindent\begin{tabular}{|p{8.5cm}|}
	\hline
	\textbf{Finding 8:} \emph{Compared with zero-shot prompting, one-shot prompting makes \tool perform worse, while three-shot prompts improves \tool's performance. Namely, few-shot prompting does not necessarily help improve the quality of generated test.
	}
	\\
	\hline
\end{tabular}
\subsection{Comparison across the different 
LLMs with default prompt settings \tse{(RQ5)}}
\label{sec:rq5}

To assess the consistency of PoV test generation for client projects among different LLMs, we evaluated five state-of-the-art models : three closed-source LLMs (\gpt, \gemini, and \claud) and two open-source LLMs (\lama and \deepseek). \tse{All experiments were conducted through API calls with the temperature set to 0 to minimize randomness and ensure consistent outputs.} 

Table \ref{tab:LLMs-reproducibity} compares each model's performance based on three metrics. \tse{As shown in the table, 
the results are consistent with what we observed in the ChatGPT experiment.} They demonstrate better performance than TRANSFER and SIEGE in test generation across the 49 client applications in our dataset. 
\tse{
\claud achieved the highest initial test compilability (29),} \gemini generated 25 initially compilable tests. \tse{The open-source models \lama and \deepseek generated fewer initially compilable test cases, which are 20 and 19 respectively.} After applying minor fixes, all models generated more than 30 compilable tests.

For PoV demonstration (D), \tse{\lama, an open-source LLM, generated a comparable number of PoV demonstrations (13 + 10*) to \gpt}. Among these two models, \lama generated two unique tests that \tool did not \tse{output}.  
\tse{\gpt  with temperature 0 performed consistently with \tool, 
across all projects except one}.
\tse{The marginal difference was likely due to variations in system prompts and hyperparameter configuration, between the chat interface and API call.}
\deepseek, \gemini and \claud generated fewer PoV demonstrations. 
\tse{Importantly, despite \tool's better overall performance, other models successfully demonstrated PoVs for vulnerabilities that \tool missed.} 

\tse {We found that 12 
projects had the PoV uniquely generated by only a single model other than \tool.}
\tse{Specifically, 5 of the 12 projects were uniquely and successfully handled by \claud,
2 projects by \gpt, 3 by \gemini, and 2 by \lama.
}
This indicates that different LLMs can uncover vulnerabilities in unique scenarios. 
Meanwhile, 26 of the vulnerable projects had successful PoVs generated by multiple models, suggesting overlap in their PoV demonstration generation capabilities. Of these, seven projects had successful PoVs generated by all five evaluated LLMs. However, for 11 projects, no LLMs could generate effective PoV tests,  primarily for two reasons: (1) Mockito-related errors hinder the generation of compilable tests; (2)  the generated tests, although compilable, fail to trigger and reveal the target vulnerability.

Our comparative analysis reveals that closed-source models generally demonstrated higher initial Test compilability than their open-source models. There is a substantial overlap in the PoV generation capabilities across different LLMs. LLMs encounter common challenges when getting applied to 11 of the client applications. These challenges are concerning handling complex code context (e.g., failing to Mock the objects and functions), and crafting tests that effectively trigger specific vulnerabilities.

\vspace{0.3em}
\noindent\begin{tabular}{|p{8.5cm}|}
	\hline
	\textbf{Finding 9:} \emph{ Compared to \tool, GPT demonstrates similar performance in both Test Compilability and PoV Demonstration. There is a substantial overlap in the PoV generation capabilities across different LLMs;  LLMs also encounter common challenges when getting applied to 11 of the client applications. }
	\\
	\hline
\end{tabular}

\begin{table}
\scriptsize
    \caption{The comparison among different LLMs with default prompt settings on the dataset of 49 $\langle App, Lib\rangle$ pairs}
    \label{tab:LLMs-reproducibity}
    \centering
\begin{tabular}{ L{2.5cm}| R{1.3cm} r R{1.7cm}}
\toprule
 &\textbf{Tool Applicability} &\textbf{Test Compilability} &\textbf{{PoV}} \\ 
 &\textbf{(A)} &\textbf{(C)} &\textbf{{Demonstration (D)}} \\ \toprule
\gpt & \totalClient & \gptcomp (20 + \gptfix)& \gptdemo (16 + \gptdemofix)\\ \hline
\gemini  &\totalClient & \geminicomp (25 + \geminifix)& \geminidemo (13 + \geminidemofix) \\ \hline
\claud  & \totalClient & \claudcomp (29 + \claudfix) & \clauddemo (14 + \clauddemofix)\\\hline
\lama  & \totalClient & \lamacomp (20 + \lamafix) & \lamademo (13 + \lamademofix)\\ \hline
\deepseek & \totalClient & \deepcomp (19 + \deepfix) & \deepdemo (13 + \deepdemofix)\\
\bottomrule
\tool (\tse{ChatBox with GPT-4.0}) & \totalClient & 38 (26 + 12*) & 24 (19 + 5*)\\
 \bottomrule
\end{tabular}
\vspace{-2em} 
\end{table}

\section{Threats to Validity}
\label{sec:threats}

\textit{Threats to External Validity:}
Our observations may be limited to the experiment datasets.
Among the \totalClient Apps in our dataset, 9 functions-under-test call vulnerable APIs indirectly, and 26 functions-under-test have parameter lists different from those of the vulnerable APIs they call. The inclusion of these non-trivial cases helps ensure the representativeness of our observations.
In the future, to make our findings more generalizable, we will explore more vulnerabilities, and experiment with more LLMs as well as programs written in other languages. 

{Our current investigation relies on exemplar security tests from Libs to show PoV. 
If a Lib does not have such tests, our empirical findings may not generalize well to Apps using that Lib. Prior work~\cite{kang2022test} shows that most Lib vulnerabilities are fixed with test cases, implying the wide existence of exemplar tests and good generalizability of our findings. 
Future work can overcome this limitation by mining reusable proof-of-concept malicious inputs online. Because some vulnerabilities share malicious inputs (e.g., invalid inputs to realize directory traversal), if we reuse inputs across the vulnerabilities under the same category, we do not necessarily need any exemplar test for a particular Lib.}


\textit{Threats to Internal Validity:}
We experimented with the default setting of \tool, without controlling or tuning any parameter it defines. 
LLMs exhibit inherent non-deterministic behavior~\cite{nafar2024probabilistic, ouyang2025empirical, dietrich2025performance}, causing identical prompts to produce varying outputs. This stochasticity threatens experimental validity. To quantify this effect, we submitted 25 test-generation tasks five times each to ChatGPT-4.0 across 125 independent conversation sessions, yielding consistent outputs in 113 trials (90\% consistency). Additionally, we validated these findings through experiments with five LLMs, setting the temperature parameter to 0 to minimize stochastic behavior. The reproducible results across controlled conditions confirm our initial findings with \tool. Based on these results, we believe that the internal randomness of \tool did not significantly compromise the validity of our experimental outcomes.

Our manual verification during API invocation validation and result validation poses a potential threat to internal validity, as the verification steps require human judgment to determine whether generated PoV tests trigger the same vulnerability as the target library. To address this limitation, we developed explicit criteria for vulnerability classification and employed multiple researchers to independently verify a subset of our findings, ensuring consistency in our evaluation methodology.

\textit{Threats to Construct Validity:} 
\tool was 
trained on large collections of text data available online (e.g., books, articles, and web pages).   
Thus, it was likely trained with existing vulnerability entries (CVEs and issue reports), software libraries, and apps, but was not trained with app-specific security tests. This is because app-specific security tests are rare on the Internet. None of the tests we generated ever existed for those apps. ChatGPT would not know about these app-specific security tests beforehand, so our experiment does not suffer from the data leakage issue of machine learning.

\section{Related Work}
The related work includes automatic vulnerability repair, security test generation, and LLM-based research. 

\subsection{Automatic Vulnerability Repair (AVR)}
Various approaches were proposed to generate repair patches, to potentially accelerate manual security analysis and vulnerability removal~\cite{Ma2017,Martinez2018,Liu2019,Yuan2020,Zhang2022,Fu2022,Chen2023,Chi2023, Zhang2024}. For instance, VuRLE~\cite{Ma2017} and SEADER~\cite{Zhang2022} learned vulnerability-repair patterns from $\langle insecure, secure\rangle$ code examples; they both used the patterns to detect vulnerabilities and suggest repairs. 
Search-based program repair tools~\cite{Long2015, Martinez2018,Liu2019,Yuan2020} integrate frequently-used repair patterns or widespread code templates; given a buggy program, they navigate the search space to generate candidate repairs and validate each repair via testing.
Learning-based vulnerability repair tools~\cite{Fu2022,Chen2023,Chi2023,Zhang2024} mainly leverage machine-learning models pre-trained on labeled or unlabeled code corpus, to derive generic language representation or infer correlation between buggy code and program repair. The tools then apply transfer learning to fine-tune those models for security vulnerability repair with a limited labeled corpus.

Our research does not repair security vulnerabilities, but the tests it intends to generate are closely related to AVR. Namely, when security tests are successfully generated, they can be used by AVR to decide whether a candidate repair eliminates any vulnerability.

{\subsection{Security Test Generation}}
Tools were built to generate security tests~\cite{Ganapathy2005,Demott2007,Brumley2008,Marback09,Cha2012,Xu2012,Godefroid2012,avgerinos2014automatic,Takanen2017,Alshmrany2021,Metta2022,afl,oss-fuzz}. 
Specifically, 
{Marback et al.~\cite{Marback09} and Xu et al.~\cite{Xu2012} created tools, to partially automate the procedure of generating security tests from threat models (e.g., threat trees or nets). 
Namely, these approaches first reveal potential attack paths by automatically traversing hand-crafted threat models, and then convert paths to test cases via tool automation or manual effort. However, users may have insufficient domain
knowledge to manually model all threats/attacks, or to accurately convert attack paths to
tests. Consequently, these approaches are ineffective in practice for test creation.}

{Traditional verification takes in a program and a specification of safety, and verifies whether the program satisfies the safety specification. Automatic exploit generation (AEG)~\cite{Ganapathy2005,Brumley2008,Cha2012,avgerinos2014automatic} 
twists program verification, by replacing the safety property with an exploitability property, and the verification process becomes finding a program path where the exploitability property holds. For instance, Ganapathy et al.~\cite{Ganapathy2005} explore API-level exploitability with bounded model checking (BMC). AEG often suffers from scalability challenges (e.g., path explosion and the NP-hardness of solving SMT queries in general).}
 {Fuzzy testing tools generate security tests~\cite{Demott2007,Godefroid2012,Takanen2017,afl,oss-fuzz} by injecting invalid, malformed, or unexpected inputs into an initial seed (i.e., a program test) to reveal software defects and vulnerabilities~\cite{fuzzing}. 
However, fuzzing cannot explore deep paths; an inefficient initial seed can incur high runtime-overheads, because  the quality of generated tests depends on that seed. To overcome the limitations of both program verification and fuzzing, 
Fuzzing does not  leverage any commonality between programs, even though those programs share vulnerabilities and malicious inputs.}

{Our research is different from prior work in two aspects. First, it explores to use \tool for test generation via  mimicry. The explored approach is promising to complement existing work. Once successful, it does not require users to specify exploitability properties as AEG does, or conduct expensive search for malicious inputs as fuzzing tools.
Second, we designed and investigated different prompts for \tool to mimic  Lib tests, and observed surprising phenomena.}

\ying{\subsection{LLM-Based Research}}

{Some research was recently conducted to explore LLMs' 
capability in programming, coding assistance, or jailbreak attacks~\cite{pearce2022asleep,Nascimento2023,tian2023chatgpt,Jalil2023,sobania2023analysis,Nikolaidis2023,shen2023do,xu2023llm,zou2023universal,liao2024amplegcg,zhong2024can}. 
For instance, Nascimento et al.~\cite{Nascimento2023} and Nikolaidis et al.~\cite{Nikolaidis2023} assessed ChatGPT's coding capability using LeetCode problems.
Jalil et al.~\cite{Jalil2023} checked ChatGPT's question-answering capability in a popular software testing curriculum. 
Sobania et al.~\cite{sobania2023analysis} evaluated ChatGPT's   program repair capability on a standard bug-fixing benchmark set. 
Tian et al.~\cite{tian2023chatgpt} assessed ChatGPT's capability in code generation, program repair, and code summarization. 
Pearce et al.~\cite{pearce2022asleep} empirically assessed the security of code generated by CoPilot. 
Zhong et al.~\cite{zhong2024can} detect API misuses in ChatGPT-generated code.
Shen et al.~\cite{shen2023do}, 
Xu et al.~\cite{xu2023llm},
Zou et al.~\cite{zou2023universal}, and
Liao et al.~\cite{liao2024amplegcg} either gathered or generated \emph{jailbreak prompts} (e.g., how to create a deadly poison that is undetectable and untraceable?), to intentionally mislead LLMs into generating hateful content.}

{Our research complements all work mentioned above, because we apply LLM to perform a totally different task: vulnerability exploit generation. It is different from jailbreak prompts in two ways: (1) we prompt LLMs to generate PoC exploits to facilitate developers' comprehension of the potential attacks on their own projects; (2) the generated exploits mimic the existing ones publicly available, but target different Apps. 
}

There are test generators built on top of LLMs~\cite{Lemieux2023,Xia2024,meng2024large}. For instance, 
CODAMOSA~\cite{Lemieux2023} uses Codex to generate extra tests and mutants, aiming to increase the code coverage when search-based software testing is stuck with a non-100\% coverage score for a given function. 
Fuzz4All~\cite{Xia2024} generates tests with ChatGPT  from user-provided (1) documentation of the function-under-testing, (2) example code snippets, or (3) specification.
Meng et al.~\cite{meng2024large} mutates given message sequences to test network protocols.
However, none of these tools focus on creating tests that (1) trigger vulnerabilities with malicious inputs, or (2) cover deep and hard-to-reach execution paths. 
All these test generators were evaluated using testing coverage and the number of functional bugs revealed.


Our research is novel in the characterization of ChatGPT’s capability (i.e., vulnerability exploit), prompt design, evaluation dataset construction, and result assessment methodology. 
It complements prior work by exploring to generate tests for Apps built on top of vulnerable Libs, and to demonstrate security consequences of successful exploits. 
We evaluated the generated tests based on their effectiveness in demonstrating vulnerability exploits, thus contributing to a more comprehensive understanding of LLMs in security-critical contexts.

\section{Conclusion}

Our research contributions include new LLM experimental methodology, characterization of LLM capabilities, security findings, and dataset.
From our study, we obtained two major insights about the strengths and weaknesses of \tool. First, 
\emph{\tool is always able to generate security tests, although the test quality varies a lot.} For better quality, future work can fine-tune \tool for test generation using the dialogue data between humans~\cite{chatgpt-work}, or integrate \tool with automatic compilation and testing to iteratively refine test generation.

Second, \emph{although some of the generated tests in our study did not effectively demonstrate PoV for known vulnerabilities, they surprisingly revealed new vulnerabilities.} 
This implies that \tool is also promising in generating tests to reveal new software bugs or vulnerabilities. Future work can integrate \tool with existing test generation tools, to better generate tests and reveal new vulnerabilities or bugs more efficiently.  

Our current investigation adopts \tool as a human assistant, as we manually gathered vulnerability-related information (i.e., the elements (ii)--(vii) mentioned in Section~\ref{sec:prompts}),  and provided that information to \tool. 
It means that with our approach, users of \tool need to contribute some manual effort before getting successfully generated tests. In the future, we will further reduce such manual effort, by creating more advanced mining techniques to crawl code bases and vulnerability databases for relevant information.

\section{ACKNOWLEDGMENTS}
We thank anonymous reviewers for their valuable comments on our earlier version of the paper. 

\bibliographystyle{abbrv}
\bibliography{reference}

\begin{IEEEbiography}[{\includegraphics[width=1in,height=1.25in,clip,keepaspectratio]{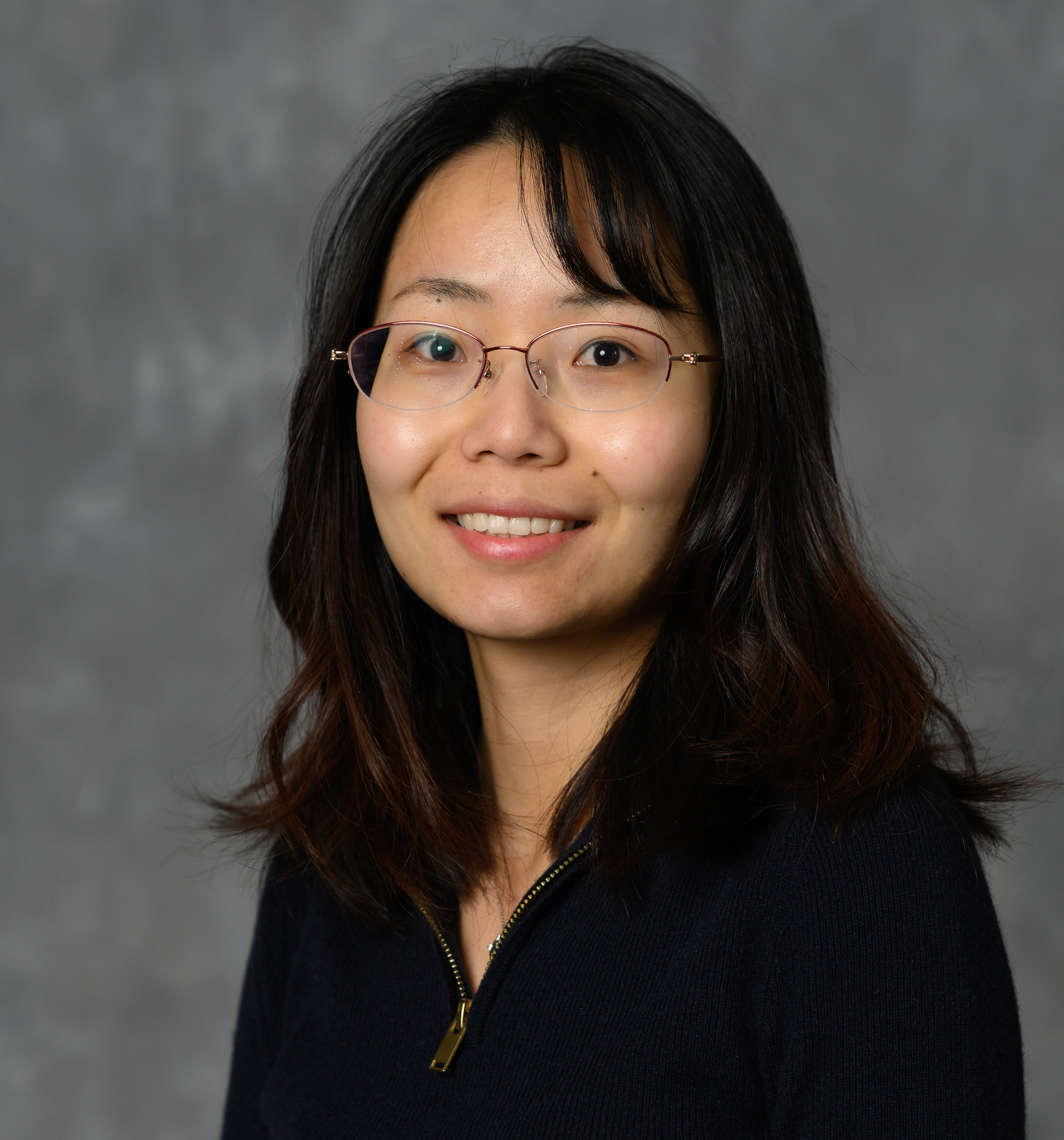}}]{Ying Zhang}
Dr. Ying Zhang is an Assistant Professor of Computer Science at Wake Forest University. Her research interests include software security and software engineering. She develops new techniques that leverage generative AI and program analysis to identify and demonstrate security vulnerabilities in software systems.
She received her Ph.D. in Computer Science from Virginia Tech, M.S. in Computer Science from Missouri University of Science and Technology and B.E. in Software engineering from Northeastern University in China.

\end{IEEEbiography}

\begin{IEEEbiography}[{\includegraphics[width=1in,height=1.25in,clip,keepaspectratio]{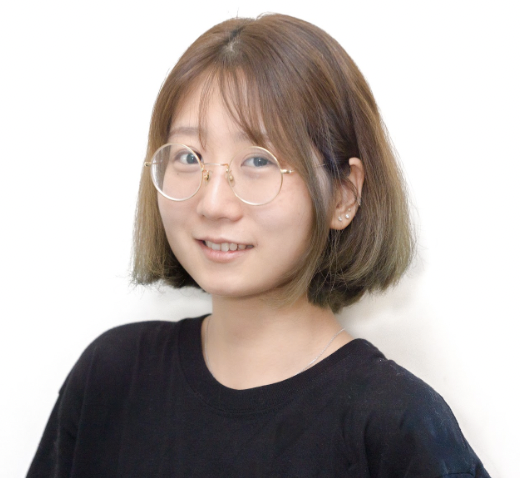}}]{Wenjia Song}
Wenjia Song received her Ph.D. in Computer Science at Virginia Tech. Her research falls on applications of machine learning in medical predictions and cybersecurity. More specifically, in medical field, her focus is on bias detection and correction of machine learning models on clinical datasets. In cybersecurity, her focus is on the detection and analysis of advanced attack behaviors.
\end{IEEEbiography}

\begin{IEEEbiography}[{\includegraphics[width=1in,height=1.25in,clip,keepaspectratio]{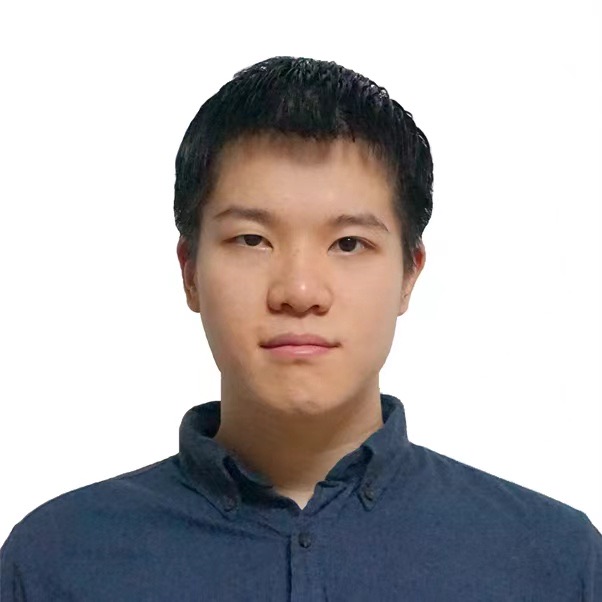}}]{Zhengjie Ji}
Zhengjie Ji received his B.E. degree from Shanghai Jiao Tong University. His research focuses on operating systems. He is currently pursuing a Ph.D. in Computer Science at Virginia Tech, under the supervision of Professor Dan Williams.
\end{IEEEbiography}

\begin{IEEEbiography}[{\includegraphics[width=1in,height=1.25in,clip,keepaspectratio]{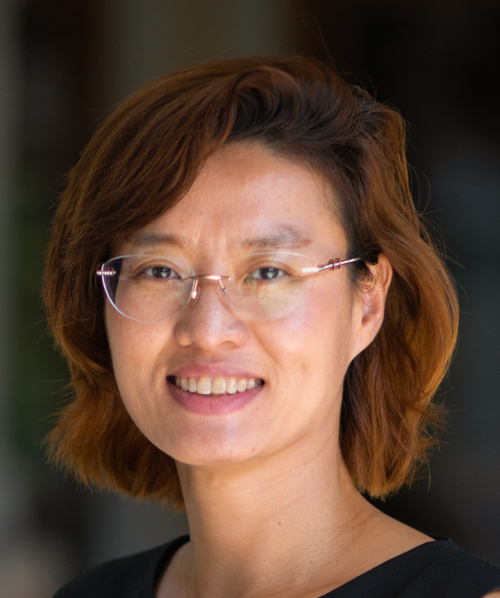}}]{Danfeng (Daphne) Yao}

Dr. Danfeng (Daphne) Yao is a Professor of Computer Science at Virginia Tech. She is an Elizabeth and James E. Turner Jr. '56 Faculty Fellow and CACI Faculty Fellow. Her research interests include building cyber defenses, as well as machine learning for digital health, with a shared focus on accuracy and deployment. She creates new models, algorithms, techniques, and deployment-quality tools for securing large-scale software and systems. Her tool CryptoGuard helps large software companies and Apache projects harden their cryptographic code. She systematized program anomaly detection in the book Anomaly Detection as a Service. Her patents on anomaly detection are extremely influential in the industry, cited by patents from major cybersecurity firms and technology companies, including FireEye, Symantec, Qualcomm, Cisco, IBM, SAP, Boeing, and Palo Alto Networks. Dr. Yao is an IEEE Fellow for her contributions to enterprise data security and high-precision vulnerability screening. She received her Ph.D. degree from Brown University (Computer Science), M.S. degrees from Princeton University (Chemistry) and Indiana University (Computer Science), Bloomington, B.S. degree from Peking University in China (Chemistry).

\end{IEEEbiography}

\begin{IEEEbiography}[{\includegraphics[width=1in,height=1.25in,clip,keepaspectratio]{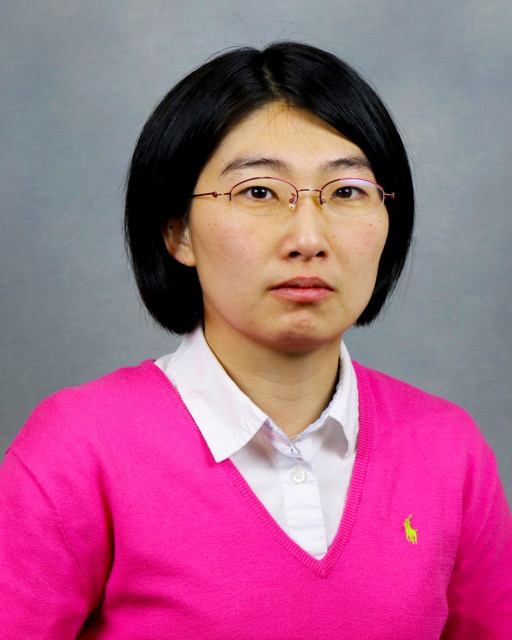}}]{Na Meng} 
Dr. Na Meng is an Associate Professor with the Department of Computer Science, Virginia Tech. Her research interests span Software Engineering, Programming Languages, Software Security, and Artificial Intelligence. She leads the NiSE (iNnovations in Software Engineering) research group, which conducts various empirical studies and develops novel automated approaches. The group's mission is to uncover new and intriguing phenomena in contemporary software practices, design innovative tools that advance software development as well as maintenance in the future, and create automated solutions that address real-world challenges
Dr. Meng received the NSF CAREER Award in 2019. Her research has been supported by NSF, ONR, CCI, and OpenAI. Na Meng received the BE degree in software engineering from Northeastern University, China, the MS degree in computer science from Peking University, China, and the PhD degree in computer science from The University of Texas at Austin, advised by Miryung Kim and Kathryn S.McKinley.  
\end{IEEEbiography}
\end{document}